\documentclass[%
 reprint,
prc,twocolumn,nofootinbib,amsmath,superscriptaddress,floatfix
 amssymb,
 aps
]{revtex4-1}

\usepackage{graphicx}
\usepackage{dcolumn}
\usepackage{bm}
\usepackage[pdftex]{hyperref}
\usepackage{wasysym}
\usepackage{amsmath}
\usepackage{mathtools}
\usepackage{xcolor}

\begin{document}
\preprint{APS/123-QED}
\title{The gamma-ray strength function of $^{89}$Y and $^{90}$Y}
\author{G.~M.~Tveten}
\affiliation{Department of Physics, University of Oslo, 0316 Oslo, Norway}
\email{g.m.tveten@fys.uio.no}

\author{T.~Renstr{\o}m}
\affiliation{Department of Physics, University of Oslo, 0316 Oslo, Norway}
\email{therese.renstrom@fys.uio.no}

\author{A.~C.~Larsen}
\affiliation{Department of Physics, University of Oslo, 0316 Oslo, Norway}

\author{H. Utsunomiya}
\affiliation{Konan University, Department of Physics, 8-9-1 Okamoto, Higashinada, Japan}

\author{K. Stopani}
\affiliation{Lomonosov Moscow State University, Skobeltsyn Institute of Nuclear Physics, 119991 Moscow, Russia}

\author{S. Belyshev}
\affiliation{Lomonosov Moscow State University, Department of Physics, 119991 Moscow, Russia}

\author{M.~Guttormsen}
\affiliation{Department of Physics, University of Oslo, 0316 Oslo, Norway}

\author{T.~Ari-izumi}
\affiliation{Konan University, Department of Physics, 8-9-1 Okamoto, Higashinada}

\author{F.~L.~Bello~Garrote}
\affiliation{Department of Physics, University of Oslo, 0316 Oslo, Norway}

\author{D.~L.~Bleuel}
\affiliation{Lawrence Livermore National Laboratory, Livermore, California 94551, USA}

\author{Y.~Byun}
\affiliation{Department of Physics and Astronomy, Ohio University, Athens, Ohio 45701, USA}

\author{T.~K.~Eriksen}
\affiliation{Department of Nuclear Physics,
Research School of Physics and Engineering,
The Australian National University,
Canberra ACT 2601,
Australia}

\author{D.~Filipescu}
\affiliation{ELI-NP, ”Horia Hulubei” National Institute for Physics and Nuclear Engineering (IFIN-HH), 30 Reactorului, 077125 Bucharest-Magurele, Romania}

\author{F.~Giacoppo}
\affiliation{Helmholtz Institute Mainz, 55099 Mainz, Germany}
\affiliation{GSI Helmholtzzentrum für Schwerionenforschung, 64291 Darmstadt, Germany}

\author{I.~Gheorghe}
\affiliation{ELI-NP, ”Horia Hulubei” National Institute for Physics and Nuclear Engineering (IFIN-HH), 30 Reactorului, 077125 Bucharest-Magurele, Romania}

\author{S.~Goriely}
\affiliation{Institut d'Astronomie et d'Astrophysique, Universit\'{e} Libre de Bruxelles, Campus de la Plaine, CP-226, 1050 Brussels, Belgium}

\author{A.~G{\"o}rgen}
\affiliation{Department of Physics, University of Oslo, N-0316 Oslo, Norway}

\author{S.~Harissopulos}
\affiliation{Institute of Nuclear Physics, NCSR “Demokritos”, Athens, Greece}

\author{S.~Katayama}
\affiliation{Konan University, Department of Physics, 8-9-1 Okamoto, Higashinada}

\author{M.~Klintefjord}
\affiliation{Department of Physics, University of Oslo, N-0316 Oslo, Norway}

\author{W.~Luo}
\affiliation{School of Nuclear Science and Technology, University of South China, Hengyang 421001, China}

\author{Y.-W.~Lui}
\affiliation{Cyclotron Institute, Texas A\& M University, College Station, Texas 77843, USA}

\author{E.~Sahin}
\affiliation{Department of Physics, University of Oslo, N-0316 Oslo, Norway}

\author{R.~Schwengner}
\affiliation{Helmholtz-Zentrum Dresden-Rossendorf, 01328 Dresden, Germany}

\author{S.~Siem}
\affiliation{Department of Physics, University of Oslo, N-0316 Oslo, Norway}

\author{D.~Takenaka}
\affiliation{Konan University, Department of Physics, 8-9-1 Okamoto, Higashinada}

\author{T.~G.~Tornyi}
\affiliation{Section of Experimental Nuclear Physics, Institute for Nuclear Research,
Hungarian Academy of Sciences, Debrecen, HUNGARY}

\author{A.~V.~Voinov}
\affiliation{Department of Physics and Astronomy, Ohio University, Athens, Ohio 45701, USA}

\author{M.~Wiedeking}
\affiliation{iThemba LABS, P.O.  Box 722, 7129 Somerset West, South Africa}

\date{\today}

\begin{abstract}
In this work, we present new data on the $^{89}$Y($\gamma$,n) cross section studied with a quasi-monochromatic photon beam produced at the NewSUBARU synchrotron radiation facility in Japan contributing torwards resolving a long standing discrepancy between existing measurements of this cross section.  Results for $\gamma$-ray strength function below threshold obtained by applying the Oslo method to $^{89}$Y($p,p'\gamma$)$^{89}$Y coincidences combined with the $^{89}$Y($\gamma$,n) data this providing experimental data for the $\gamma$-ray strength function of $^{89}$Y for $\gamma$ energies in the range of $\approx 1.6$ Mev to $\approx$ 20 MeV. A low-energy enhancement is seen for $\gamma$-rays below $\approx 2.5$ MeV. Shell-model calculations indicate that this feature is caused by strong, low-energy $M1$ transitions at high excitation energies. The nuclear level density and $\gamma$-ray strength function have been extracted from $^{89}$Y($d,p \gamma$)$^{90}$Y coincidences using the Oslo method. Using the ($\gamma,n$) and ($d,p\gamma$) data as experimental constraints, we have calculated the $^{89}$Y($n,\gamma$)$^{90}$Y cross section with the TALYS reaction code. Our results have been compared with directly measured (n,$\gamma$) cross sections and evaluations. The $N=50$ isotope $^{89}$Y is an important bottleneck in the s-process and the magnitude of the $^{89}$Y(n,$\gamma)$ cross section is key to understanding how s-process stars produce heavy isotopes. 
\end{abstract}

\maketitle
\section{Introduction}
\label{sec:int}

Explaining the observed distribution of heavy element abundances in our solar system is a pressing scientific question. The quest for a quantitative understanding of nucleosynthesis involves exploring and understanding a complex interplay between nuclear properties and extreme, astrophysical environments. Our understanding of the stellar processes responsible for the production of elements heavier than iron has improved significantly since the first serious attempts at explaining stellar nucleosynthesis in the seminal works of Burbidge \textit{et al.}~\cite{burbidge1957} and Cameron~\cite{cameron1957} in the 1950’s. Despite recent advances, several open questions remain and better determined nuclear data is key to answering these open questions~\cite{Arnould2007,Rauscher2013}. 

\begin{figure}[tb]
\includegraphics[width=0.45\textwidth]{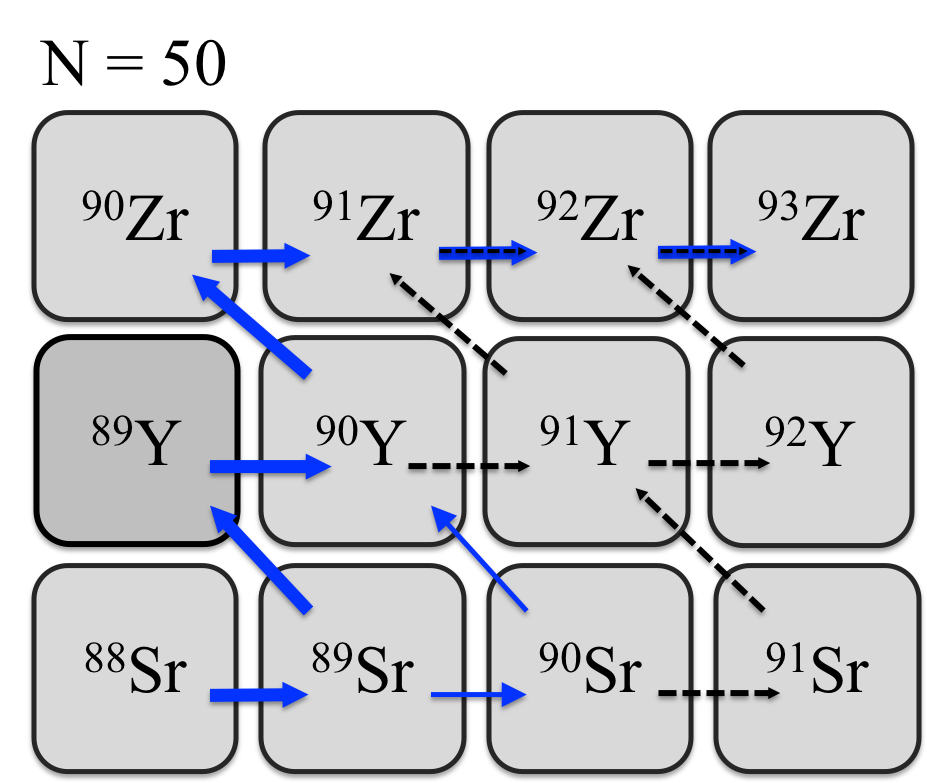}
\caption{(Color online) Part of the s-process network is shown here. The blue arrows indicate the main reactions that involve $^{89}$Y for the case of an s-process taking place in an AGB star, while the dashed arrows indicate branchings that in a more massive AGB star may impact the production of Zr isotopes. \label{fig:sflow}}
\end{figure}

Elements heavier than iron are mainly produced by the slow neutron capture process, $s$-process, that takes place in asymptotic giant branch stars or in the rapid neutron capture process, $r$-process. We recently found observational support for that this process takes place in neutron star mergers (not excluding other possible r-process sites such as core collapse supernovae)~\cite{Rauscher2013,Savchenko2017}. The isotope $^{89}$Y is produced by both the s- and r-processes. Since it is an $N=50$-isotope, the neutron capture cross section on this isotope is rather low, making it a bottleneck for the s-process and the yield of heavier isotopes. Understanding the  $^{89}$Y(n,$\gamma)^{90}$Y reaction cross section is therefore key to understanding how isotopes heavier than $^{89}$Y are formed by the s-process in the evolved stars on the asymptotic giant branch (AGB) of the Hertzsprung-Russell diagram. Reactions that involve $^{89}$Y are important to determine the total s-process production of isotopes in the mass region $78 \leq A \leq 92$ \cite{sprocessN50} and the neutron cross section of $^{89}$Y also influences the whole s-process abundance distribution~\cite{0004-637X-601-2-864}. The r-process contribution to the solar abundances, a key set of abundance data to which galactic chemical evolution model outputs is routinely compared, is commonly obtained by subtracting the s-process contribution. For this reason, s-process isotope production is key to understanding also the r-process. Furthermore, the abundance of $^{89}$Y is one of three so-called s-light abundances that are used as reference to compare theoretical models of stellar nucleosynthesis and galactic chemical evolution to abundance observations. The part of the s-process reaction network that involves $^{89,90}$Y is shown in Fig.\ref{fig:sflow}. Also note that in AGB stars more massive than typically $4~M_{\odot}$, the thermal pulses are hot enough to burn $^{22}$Ne at the bottom of the pulse leading to a rather large neutron flux (with densities of the order of $10^{12}$~cm$^{-3}$). In this case an important amount of the unstable $^{90}$Sr (with a half-life of $t_{1/2}=28.8$~y),$^{90}$Y ($t_{1/2}=2.7$~d) and $^{91}$Y ($t_{1/2}=58.5$~d) is produced. These branchings in the neutron-rich region may impact the production of Zr isotopes by bypassing $^{90}$Zr \cite{Karinkuzhi18}, but are still affected by nuclear physics uncertainties regarding the $^{90}$Sr(n,$\gamma)^{91}$Sr, $^{90}$Y(n,$\gamma)^{91}$Y and $^{91}$Y(n,$\gamma)^{92}$Y reaction rates, as also indicated in Fig.\ref{fig:sflow}. 

The reaction cross sections needed for large reaction network calculations are calculated in the statistical framework of Hauser and Feshbach~\cite{hauser}, except for cases where experimental cross sections are available. Two key ingredients for such calculations are the nuclear level density (NLD) and the $\gamma$-ray strength function ($\gamma$SF). The cross section of neutron capture on unstable isotopes is challenging to study experimentally, and benchmarking theoretical models of NLD and $\gamma$SF needed for calculating cross sections is therefore important also in the context of the s-process. 

In this work we present novel measurements of the $(\gamma,n)$ reaction cross section on $^{89}$Y made at the NewSUBARU synchrotron facility~\cite{Ando_new_sub_design, newsubweb}. The inverse Compton scattering method was used to produce $\gamma$-ray energy photon beams in the range of $S_{\textrm{n}} \leq E_{\gamma} \leq S_{2\textrm{n}}$. This photon beam was used to make neutron measurements between the neutron binding energy, $S_n=11.482$ MeV, and the two neutron separation energy, $S_{2n}=20.835$ MeV. We combine our results obtained at NewSUBARU with the $\gamma$SF measured below particle threshold for $^{89}$Y obtained using the Oslo method~\cite{unfolding,schiller2000,omsystematics}, making use of the principle of detailed balance for emission and absorption of radiation \cite{blatt1952}. The NLD of ${}^{89, 90}$Y below $S_n$ have previously been reported in Ref. \cite{guttormsen_yttrium_2014} and the $\gamma$SF of $^{89}$Y in Ref. \cite{PhysRevC.93.045810}. We focus here on the experimental details and results of the ${}^{89}$Y($d,p)^{90}$Y experiment where the NLD and $\gamma$SF of $^{90}$Y below the neutron binding energy, $S_n$, were extracted. The Y-isotopes close to stability are not expected to display substantial structure effects and therefore the experimental results for the $\gamma$SF of $^{89}$Y are used in combination with the $^{90}$Y $\gamma$SF and NLD to constrain the $^{89}$Y$(n,\gamma)^{90}$Y cross section and to calculate the Maxwellian averaged reaction rates for astrophysical relevant temperatures. 

\section{Measuring $^{89}$Y($\gamma$,n): Setup and method}
The measured photonuclear cross section for the exclusive one-neutron channel, $\sigma_{\textrm{exp}}$, for an incoming photon beam with maximum beam energy, $E_{\textrm{max}}$, is given by,
\begin{equation} \label{eq:gammancs}
\sigma_{\textrm{exp}}=\int_{S_n}^{E_{\textrm{max}}}D^{E_{\textrm{max}}}(E_{\gamma})\sigma(E_{\gamma})dE_{\gamma}=\frac{N_n}{N_tN_{\gamma}\zeta\epsilon_ng},
\end{equation}
where $D$ is the normalized energy distribution of the photon beam and $\sigma(E_{\gamma})$ is the photoneutron cross section as function of photon energy, $E_{\gamma}$. The number of neutrons detected is $N_n$, $\epsilon_n$ represents the neutron detection efficiency, $N_t$ the number of target nuclei per unit area, $N_{\gamma}$ the number of photons incident on target and $\zeta=(1-\exp^{-\mu t})/\mu t$ is a correction for the self-attenuation effect in a thick-target measurement and finally,$g$, is the fraction of photons with $E_{\gamma}>S_n$. To determine $\sigma(E_{\gamma})$ from $\sigma_{\textrm{exp}}$, we need to determine experimentally the other parameters on the right hand side of \ref{eq:gammancs} and the energy distribution of the photon beam.  

The experiment was carried out using photon beams with maximum energies energies in the range of 11.6-20.0 MeV with FWHM 0.21 - 0.68 MeV. The energy distribution of the photon beams used in this work are shown in Fig.\ref{fig:allbeams}). The beams were produced through inverse Compton scattering between Nd:YVO$_4$ laser photons ($\lambda$ = 1064 nm) and relativistic electrons at the NewSUBARU storage ring \cite{Ando_new_sub_design}. The laser Compton scattering (LCS) photons resulted in narrowly distributed, pencil-like beams. The experiment was set up at BL01, situated at the end of one of the two 14 m long straight sections of the storage ring.  Electrons are injected into the ring at $\simeq$ 1.0 GeV and can be decelerated down to $\simeq$ 0.5 MeV or accelerated up to $\simeq$ 1.5 GeV. The energy of the photon beam is varied by changing the energy of the electron beam, rather than the wavelength of the laser photons. 

\begin{figure}[tb]
\includegraphics[width=0.45\textwidth]{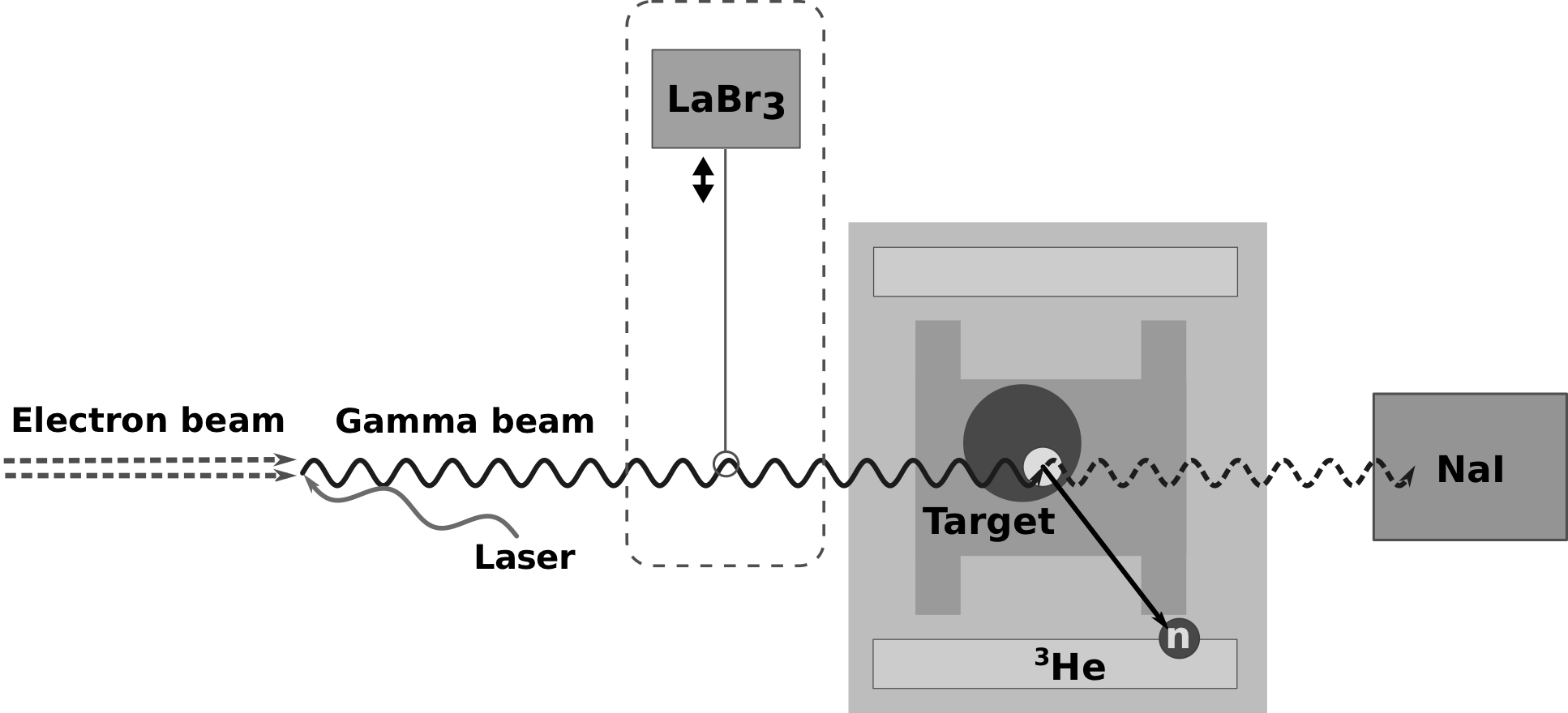}
\caption{A schematic illustration of the experimental setup, including the laser-electron collision. The angle between the laser photon and electron is added for illustration only, at NewSUBARU the laser beam approaches the electron beam head-on.\label{fig:gacko}}
\end{figure}

The photon beam was directed at a $^{89}$Y target with areal density of 1.873 g/cm$^2$. The High Efficiency Neutron Detector, based upon the ring-ratio technique developed by Berman \textit{et al.} \cite{berman1967}, was used to detect the neutrons emitted from the $(\gamma,n)$-channel. A 8"x12" NaI(Tl) scintillator detector was placed behind the target and the neutron detector directly in the beam-line to continuously monitor the number of photons per beam-bunch. For neutron detection the signals were read out from the detector using a combined amplifier and discrimination module, and the number of neutrons detected were counted with a scaler unit. The schematic layout of the setup is provided in Fig. \ref{fig:gacko}. Further details on the setup and analysis is provided in what follows.
\subsection{Determining $N_{\gamma}$ and the beam profile}
The LCS photons are produced in head-on collisions between laser photons and electrons. The energy of a photon emitted after scattering off an electron is given by the following relation
\begin{equation} \label{eq:comptonscatt}
E_{\gamma} = \frac{4{\gamma}^2E_L}{1+(\gamma \theta)^2 + 4\gamma E_L/(mc^2)}
\end{equation}
where $E_L$ is the laser photon energy, $\gamma$ is the relativistic factor $\gamma = E_e/mc^2 = 1/\sqrt{1-\beta^2}$ where $E_e$ is the incident electron energy and $mc^2 = 0.511$ MeV is the electron energy at rest, $\beta=v/c$ where $v$ is the velocity of the electrons and $c$ is the speed of light and $\theta$ is the scattering angle of the scattered photon relative to the electron beam axis. The energy spread of the photon beam is mainly due to the electron beam emittance and the angular divergence of the backscattered photon beam, the latter of which was limited by two lead collimators placed between the laser-electron interaction point and the experimental station at BL01.
The laser beam was provided by a Q-switch Nd:YVO$_4$ laser with wavelength, $\lambda=1064$ nm, and maximum power $= 35$ W. The laser was operated at the internal frequency at 20 kHz and was gated with external switching gates at 10 Hz providing a macroscopic time structure of 80 ms beam-on and 20 ms beam-off. The macroscopic time structure of the beam-off was used as gate to generate background spectra during the runs. 
  
The energy of the electron beam, and consequently the maximum energy of the photon beam, was calculated from the nominal electron energy setting of the storage ring using distinct calibration coefficients for the beam energy for nominal energies 974 MeV $\leq E_0 \leq$ 1250 MeV \cite{Shima2014} and and for 500 MeV $\leq E_0 <$ 974 MeV \cite{Uts2014}. This calibration has an accuracy in the order of $10^{-5}$. 

A total of 12 photon beam energies ranging from 11.6 MeV - 16.0 MeV were provided by decelerating the electrons and 4 photon beam energies in the range of 17.0 - 20.0 MeV were provided by accelerating the injected electrons. For comparison, $S_n=11.474$ MeV and $S_2n=20.8257$ MeV for $^{89}$Y, meaning that we probed the full range of the exclusive $(\gamma,n)$ channel. The $(\gamma$,np) channel opens at 18.190 MeV. The threshold for $(\gamma$,p) is at 7.076 MeV, but we assume throughout our analysis that any effect of this channel can be neglected since the $(\gamma,n)$ channel dominates. 

After every change of electron energy (and thus photon beam energy), the alignment of the setup was checked by inspecting the light spot produced by the synchrotron radiation from NewSUBARU, ensuring that it was centered on the center of the neutron detector and the NaI(Tl)-scintillator detector at the end of the beam-line. 

To determine the energy profile of the photon beam, a $\gamma$-spectrum was measured at the beginning and end of each run with a given energy. For this purpose a $3.5^{\prime\prime}\times 4.0^{\prime\prime}$ LaBr$_3$(Ce) detector was placed directly in front of the photon beam. To avoid pile-up in this detector, the laser power was set to a lowest setting and a 2 cm thick lead slab was placed in front of the scintillator to attenuate the beam to $\ll$ 1 photon per bunch to avoid multi-photon events. A time gate was used to take a background spectrum in parallel by gating on the laser-off time window. 

To obtain the actual energy profile of the incident photon beam, the response function of the LaBr$_3$(Ce)-scintillator detector must be taken into account. For this purpose a simulation package has been developed for simulating the response of the LaBr$_3$(Ce) detector to the photon beam using the framework of GEANT4 9.6\cite{geant1,geant2,geant3,thesisioana}. The interaction between a laser photons and electrons, as well as the transport of photons back into the experimental hall and the LaBr$_3$(Ce) scintillator detector, are simulated with the GEANT4 package. The simulations include beam-line elements such as the vacuum tube and the collimators. The emittance parameters of the electron beam are varied by hand, starting at the emittance values of Ref.\cite{Horikawa2010} and varying the emittance ellipse parameters within the typical deviations measured for the NewSUBARU storage ring, until a good agreement between the simulated spectrum and experimental spectrum has been obtained. The incident photon beam that provides a the best agreement between simulated is accepted as the energy distribution of the photon beam. This simulation procedure was repeated for each beam energy as the emittance of the electron beam will change as the energy of the beam is changed and also as function of storage time in the ring. 

\begin{figure}
\includegraphics[width=0.51\textwidth]{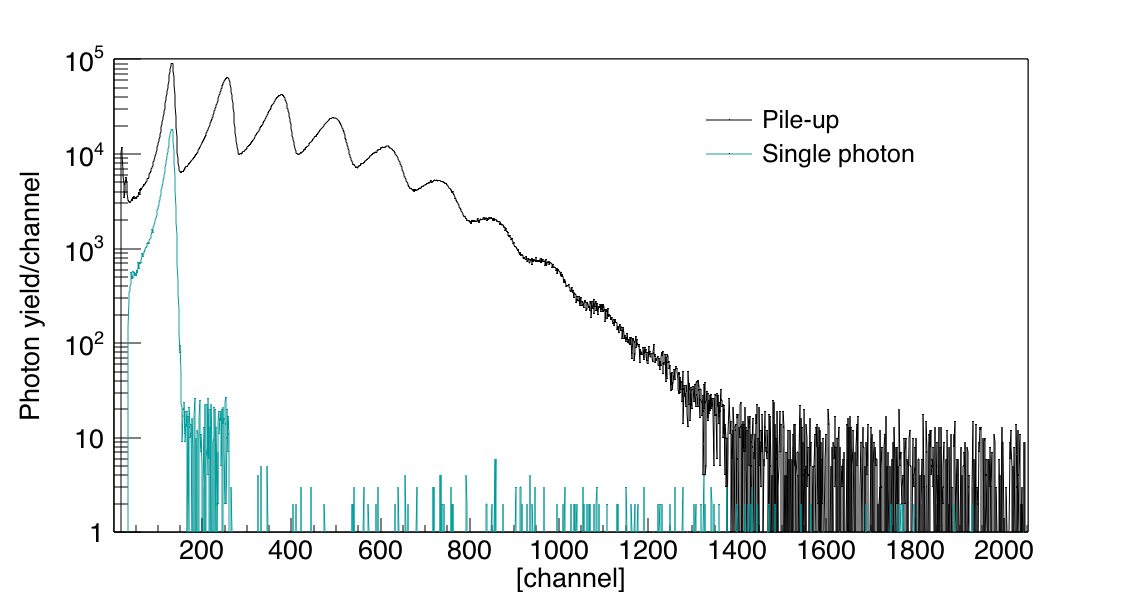}
\caption{(Color online) The pile-up and single photon spectrum for $E_{\gamma,\textrm{max}}=14.68$ MeV. In this case the average number of photons per bunch is 3.4 and $N_{\gamma} = 4.32 \times 10^7$. \label{fig:pileup}}
\end{figure}

Since both the electrons in the storage ring and the laser photons are packed in small bunches due to the microstructure of the colliding beams, the photon beam resulting from collisions is also bunched. The electron beam is bunched at 500 MHz with a bunch width of 60 ps and the laser beam at 20 kHz with a bunch width of 60 ns. Consequently, the photon beam resulting from collisions has the same bunch-structure as that of the laser photons. The photons passing through the target without interacting are detected in the NaI(Tl) detector behind the neutron detector. Since photons within a given 60 ns bunch cannot be resolved in time by the NaI(Tl) scintillation detector, the signals pile up, generating a pile-up (or multi-photon) spectrum. From the shape of the measured pile-up spectrum, the mean number of photons per bunch was deduced, and consequently the total number of photons, $N_{\gamma}$, could be calculated \cite{Kondo2011} according to the following equation
\begin{equation}
N_{\gamma} = \frac{\langle ch \rangle_{\textrm{pile-up}}}{\langle ch \rangle_{\textrm{single}}}(\sum n_i)_{\textrm{pile-up}}
\end{equation}
where $\langle ch \rangle_{\textrm{pile-up}}$ is the mean channel of the pile-up spectrum, $\langle ch \rangle_{\textrm{single}}$ the mean channel of the single photon spectrum and $(\sum n_i)_{\textrm{pile-up}}$ the total number of counts, $n_i$, for all channels $i$. A typical single photon and pile-up spectrum is shown in Fig. \ref{fig:pileup}. 
In a recent work by Utsunomiya \textit{et al}, submitted to Nucl. Instrum Meth. A, the experimental formula used to determine the mean number was thoroughly investigated with the Poisson-fitting method. It was shown that the inherent uncertainty of this method for determining photon flux determined from the pile-up spectrum (provided that the spectra are free from quenching effects of the photomultiplier tube of the NaI(Tl) detector as is the case here) is less than 0.1\%. The main contribution to the uncertainty of the pile-up technique is consequently related to experimental conditions leading to ambiguity in what channel to set the lower threshold for analysis at and where to cut off the small 2-photon contribution in the single-photon spectrum. This uncertainty has been estimated to be $\approx$ 1\%.

\subsection{Neutron detection}
\begin{figure}[h]
\includegraphics[width=0.49\textwidth]{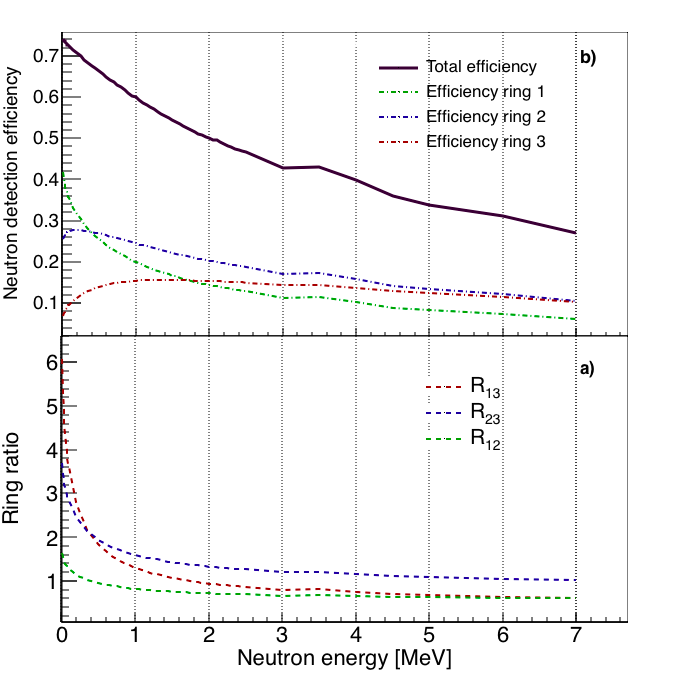}
\caption{(Color online) a) The ring ratio curves for the High Efficiency Neutron Detector and b) the efficiency curves, as function of the neutron energy, for the three rings and th total efficiency. \label{fig:thecurves}}
\end{figure}

As mentioned previously, the neutrons were detected with the high-efficiency 4$\pi$ neutron detector consisting of 20 $^3$He-filled proportional counters embedded in a polyethylene moderator of $36 \times 36 \times 50$ cm$^3$ fully covered by a neutron absorbing material in order to reduce the neutron background. The proportional counters were arranged in three rings of 4, 8, and 8 $^3$He counters placed at distances of 3.8 (ring 1), 7.0 (ring 2), and 10.0 cm (ring 3), respectively from the photon beam axis \cite{Filipescu2014}. The average neutron energy was determined by the ring-ratio technique originally developed by Berman \textit{et al.} \cite{berman1967}, where differential moderation provides a measure of the average energy of the detected neutrons. The discriminator threshold of the counters was adjusted to exclude signals from X-ray and $\gamma$, consequently only counting signals originating from the $^3$He(n,p)$^3$H reaction in the counters. By taking the ratios of counts for detectors in the different rings of detectors, $R_{12}$, $R_{23}$ and $R_{13}$, the average energy of the emitted neutrons was determined. The ring-ratio curves (see Fig. \ref{fig:thecurves}) used to determine the average neutron energy, and thus the detection efficiency, were determined by simulating the response of the detector using monochromatic neutron sources \cite{neutrondet}. 

\begin{figure}[h]
\includegraphics[width=0.49\textwidth]{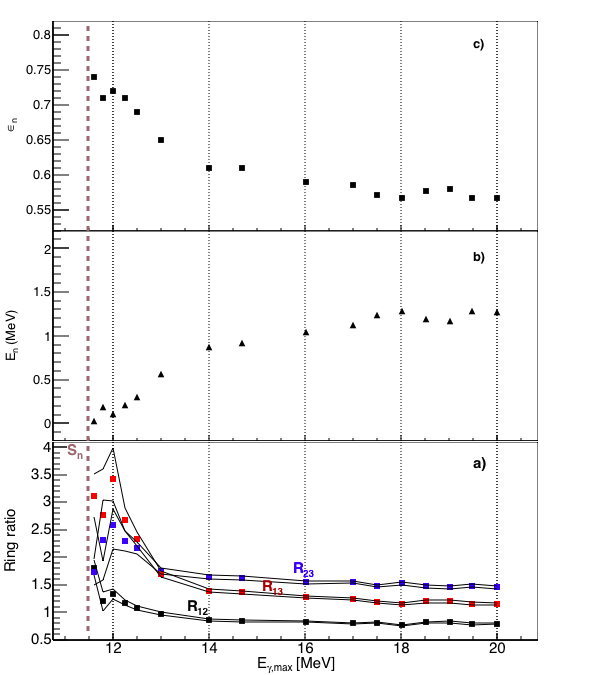}
\caption{(Color online) a) The ring ratios determined for each photon beam energy used in the current experiment and b) the average values for the neutron energy, $E_n$, used to determine the c) total detection efficiencies, $\epsilon_{n}$, used to determine the total number of neutrons emitted. The values in b) and c) are given without error bars. See the discussion in Sec.\ref{sec:errorprop} for details on uncertainties and error propagation. \label{fig:neutroneff}}
\end{figure}

The total neutron detection efficiency is $> 60\%$ for neutrons with energies less than 1 MeV. The detection energies for the neutrons detected in this experiment are shown in panel c of Fig.\ref{fig:neutroneff}. The neutron detection efficiencies of the three rings were recently remeasured using a calibrated $^{252}$Cf source with an emission rate of $2. 27 \times 10^4$ s$^{− 1}$ with 2.2$\%$ uncertainty at the National Meteorology Institute of Japan \cite{Nyhus2015}. Details about the neutron detector can be found in Ref. \cite{neutrondet}. The target sample was kept in a aluminum holder shaped as a cylinder placed at the center of the neutron detector setup. The ring ratios obtained in this experiment are shown in panel a of Fig. \ref{fig:neutroneff} and the corresponding average neutron energies (shown in panel b) were determined using the curves shown in Fig. \ref{fig:thecurves}. 

\subsection{Error propagation}\label{sec:errorprop}
In this work we have performed error propagation
analysis by Monte Carlo sampling. The starting point for determining the neutron detection efficiency is the ring ratios. In the Monte Carlo simulations, the measured number of detected neutrons for each ring, both on gate $N_{1,ON}$, $N_{2,ON}$, $N_{3,ON}$ and off gate $N_{1,OFF}$, $N_{2,OFF}$, $N_{3,OFF}$, were varied. The number in the subscript stands for the ring number. We assumed that the measured values to vary as a Gaussian distribution where the mean value $\mu_{i,A}$ was taken to be the originally measured values, $\mu_{i,A}=N_{i,A}$ and the standard deviation, $\sigma_{i,A}$, was taken to be $\sigma_{i,A}=\sqrt{N_{i,A}}$, where $i$ is the ring number and $A$ is ON or OFF. For each sampling repetition, the three ring ratios, $R_{i,j}=\frac{N_i}{N_j}, i\neq j$, were calculated, where $N_i$ is the number of neutrons after background subtraction. The average neutron energy, $E_n$, was then calculated and the neutron detection efficiency, $\epsilon_n$, by accessing the ring-ratio-curves for the neutron detector.

The ring-ratio-curve itself has an uncertainty stemming from the uncertainty in absolute calibration of the efficiency of the detector. This uncertainty was assumed to be the same as the uncertainty of the main calibration point of the efficiency curve and the whole efficiency of the detector was sampled independently from a Gaussian distribution where the mean value was taken to be $\epsilon_n$ and the standard deviation to be $0.022 \epsilon_n$. 

The total uncertainty of $N_{\gamma}$ is taken to be $\approx$ 1.0$\%$ in this work, as the 1.0\%. The errors of $N_{\gamma}$ are also assumed to be distributed according to a Gaussian with the mean $\mu=N_{\gamma}$. The number of photons were also sampled independently from the other variables. We did not attempt to quantify the uncertainty of the energy profile determined through GEANT4-simulations. As this error is expected to be small, but the work would be detailed and rather technical, such an investigations is left to be carried out in future work. 

Finally, the deviation for each run was determined by fitting a Gaussian function to the simulated distribution of cross section values resulting in the standard errors, SE$_{CS}$, of each run. The results are presented in Tab. \ref{tab:errortab}. As one would expect and has been reported in earlier works, see e.g. Ref.\cite{neutrondet}, the largest errors occur for the photon beam energies closest to $S_n$. This is mainly due to the low statistics in the neutron number due to  the low $(\gamma,$n)-cross section close to particle threshold. The $\pm$ 1 $\sigma$ limits of the unfolded cross section shown in Fig.\ref{fig:gnfinal}, $\sigma(E_{\gamma})$, were finally obtained by unfolding the monochromatic cross section, $\sigma_{EXP} \pm 1 \textrm{SE}$, where $\textrm{SE}$ is the standard error obtained in the Monte Carlo simulations. The values are also provided in Table \ref{tab:errortab}.
\begin{table}[tb]
\centering
%
\caption{The simulated errors of the measured cross sections for the 17 runs, before any corrections for the energy distribution of the photon beam. $E_{nom}$ is the nominal beam energy, $\sigma_{CS}$ the average cross section for the full photon beam distribution and SE$_{CS}$ the standard error of the cross section.  See the text for details on the simulations.}
\label{tab:errortab}
\begin{tabular}{llll}
\hline
$E_{nom}$ [MeV]& $\sigma_{\textrm{exp}}$ [mb] & SE$_{CS}$ [mb]  & SE$_{CS}$ [\%]  \\
\hline
801  & 0.60   & $2.9\cdot10^{-2}$ & 4.8 \\
808  & 0.86   & $6.4\cdot10^{-2}$ & 7.5 \\
808  & 0.88   & $8.0\cdot10^{-2}$ & 9.1 \\
815  & 3.5	  & $1.7\cdot10^{-1}$ & 4.9 \\
824  & 7.6    & $2.6\cdot10^{-1}$ & 3.5 \\
832  & 12.3   & $3.1\cdot10^{-1}$ & 3.1 \\
849  & 20.6   & $5.8\cdot10^{-1}$ & 2.8 \\
882  & 41.3   & 1.1               & 2.7 \\
904  & 62.0   & 1.6               & 2.7 \\
946  & 126.4  & 3.3               & 2.6 \\
976  & 165.3  & 4.2               & 2.6 \\
991  & 172.8  & 4.4               & 2.6 \\
1006 & 156.3  & 4.2               & 2.7 \\
1020 & 143.5  & 3.8               & 2.7 \\
1034 &125.1   & 3.4               & 2.7 \\
1047 & 116.5  & 3.2               & 2.7 \\
1061 & 106.0  & 2.9               & 2.8 \\
\hline
\end{tabular}
\end{table}


\subsection{Correction for photon beam energy profile}
\begin{figure}[h]
\includegraphics[width=0.45\textwidth]{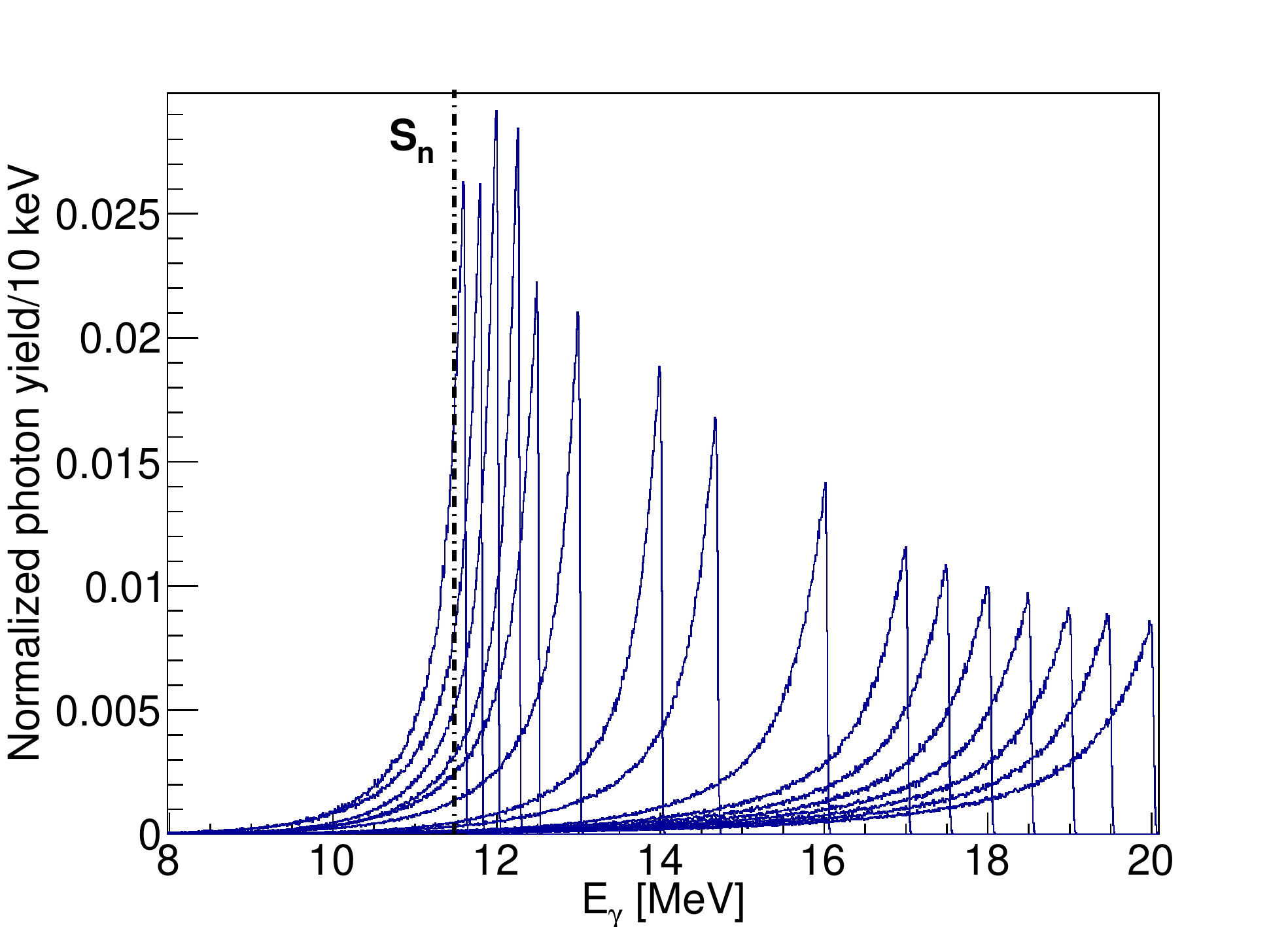}
\caption{The incident photon beam energy distributions for all runs, as determined by GEANT4 simulations. The total area of the beam profiles have been normalized to 1.\label{fig:allbeams}}
\end{figure}

A first approximation for the cross section can be obtained by using the maximum photon beam energy , $E_{\gamma,\textrm{max},i} $ for a given $E_e,i$ and assuming that the photon energy is monochromatic. These values are provided in Table \ref{tab:errortab}. This measured quantity that we from now on call $\sigma_{\textrm{exp}}$, measured for $E_{\gamma,\textrm{max}, i}$, represents the integrated cross section for the whole range of photon beam energies from $S_n-E_{\gamma,\textrm{max}, i}$. To obtain $\sigma(E_{\gamma})$, the energy profile must be taken into account. The photon beam profiles for this experiment, as determined using GEANT4 simulations, are shown in Fig. \ref{fig:allbeams}. 
A recently developed procedure has been used to determine the cross section as a function of photon energy, $\sigma(E_{\gamma})$, by unfolding with the simulated beam profiles from our integrated cross section values measured for each photon beam energy, $\sigma_f$
\begin{equation}
\sigma_f = \int_{S_n}^{E_{\gamma, \textrm{max}}} n(E_{\gamma})\sigma(E_{\gamma})dE_{\gamma}
\end{equation}
where $n(E_{\gamma})$ is constituted of our simulated photon beam profiles. The unfolding method we have developed is inspired by the well tested unfolding method developed for the Oslo method \cite{unfolding}.
We formulated the problem as a set of linear equations
\begin{equation}
\sigma_{\rm f }=\bf{D}\sigma,
\end{equation}
where the indexes $i$ and $j$ of the matrix element $D_{i,j}$ corresponds to $E_{\gamma,\rm max}$ and $E_{\gamma}$, respectively. The $D_{i,j}$ elements are non-zero for $j$ corresponding to $E_{\gamma}$ values
fulfilling the condition $ S_n - \Delta  \leq  E_{\gamma} \leq E_{\rm max} + \Delta$, where $\Delta$ is the resolution of the full-energy peak. 
Thus, the set of equations is given by
\begin{equation}
\begin{pmatrix}\sigma_{\rm{1}}\\\sigma_{\rm{2}}\\ \vdots \\ \sigma_N \end{pmatrix}_{\rm f}\\\mbox{}=
\begin{pmatrix}D_{ 11} & D_{ 12}& \cdots &\cdots &D_{ 1M} \\ D_{ 21} & D_{ 22}&
\cdots & \cdots &D_{ 2M} \\ \vdots &\vdots & \vdots & \vdots & \vdots \\ D_{ N1} & D_{ N2}& \cdots & \cdots &D_{ NM}\end{pmatrix}
\begin{pmatrix}\sigma_{1}\\\sigma_{2}\\ \vdots \\ \vdots \\\sigma_{M} \end{pmatrix}.
\label{eq:matrise_unfolding}
\end{equation}
Each row of $\bf{D}$ corresponds to a GEANT4 simulated photon
beam profile belonging to a specific measurement characterized by $E_{\gamma,\rm max,i}$.
In this experiment, we measured $N=16$ beam energies, but the beam profile is simulated with $M = 2000$ energy bins. The system of linear equations in Eq.~(\ref{eq:matrise_unfolding})is underdetermined and the $\sigma(E_{\gamma})$ cannot be determined by matrix inversion. In order to find $\sigma(E_{\gamma})$,
we utilize the following iterative algorithm to unfold for the photon beam profile:
\begin{itemize}
\item [1)] As a starting point, we choose for the 0th iteration, a constant trial function $\sigma^0$.
This initial vector
is multiplied with $\bf{D}$ and we get the 0th folded vector $\sigma^0_{\rm f}= {\bf D} \sigma^{0}$.
\item[2)] The next trial input function, $\sigma^1$, can be established by adding the difference of
the experimentally measured spectrum $\sigma_{\rm{exp}}$ and the folded spectrum $\sigma^0 _{\rm f}$,
to $\sigma^0$. In order to be able to add the folded and the input vector together, we first perform a spline
fit on the folded vector, then interpolate with a cubic spline, so that the two vectors have equal dimensions. Our new input vector is:

\begin{equation}
\sigma^1 = \sigma^0 + (\sigma_{\rm{exp}}-\sigma^0 _{\rm f}).
\end{equation} 

\item[3)] The steps 1) and 2) are iterated $i$ times giving
\begin{eqnarray}
\sigma^i_{\rm f} &=& {\bf D} \sigma^{i}
\\
\sigma^{i+1}     &=& \sigma^i + (\sigma_{\rm{exp}}-\sigma^i _{\rm f})
\end{eqnarray}
until convergence is achieved. This means that
$\sigma^{i+1}_{\rm f} \approx \sigma_{\rm exp}$ within the statistical errors.
In order to quantitatively check convergence, we calculate the reduced $\chi^2$ of $\sigma^{i+1}_{\rm f}$ and
$\sigma_{\rm{exp}}$ after each iteration.
\item[4)] Finally, an energy dependent smoothing was applied. No structures finer than the full width half maximum of the photon beam may be expected to be resolved and are thus removed by smoothing. 
\end{itemize}
In this work we needed 6 iterations to obtain convergence within the statistical uncertainties. The final result (Table \ref{tab:gntab}) is shown in Sec. \ref{sec:data} Fig.\ref{fig:gnfinal} where we compare with ($\gamma$,n) cross section data from previous works by Berman et al. \cite{berman1967} and Lepretre et al. \cite{lepretre1971}. 
\begin{table}[tb]
\centering
%
\caption{The final, unfolded cross section $\sigma$ evaluated at the maximum photon beam energy, $E_{\gamma,\textrm{max}}$, of each run, and the standard error, SE$_{CS}$, of the unfolded cross section also evaluated at $E_{\gamma,\textrm{max}}$. The number of digits for $E_{\gamma, \textrm{max}}$ indicates how well the beam profile has been determined.}
\label{tab:gntab}
\begin{tabular}{lll}
\hline
$E_{\gamma, \textrm{max}}$ [MeV]& $\sigma$ [mb] & SE$_{CS}$ [mb]\\
\hline
    11.80  &    3.1  &    0.2 \\
    12.00  &    8.4  &    0.4 \\
    12.26  &   15.7  &    0.5 \\
    12.50  &   24.5  &    0.8 \\
    13.00  &   30.4  &    0.9 \\
    14.00  &   59.5  &    1.6 \\
    14.68  &   90.4  &    2.4 \\
    16.02  &  177.8  &    4.6 \\
    17.00  &  214.9  &    5.6 \\
    17.50  &  191.2  &    5.0 \\
    18.02  &  139.5  &    3.8 \\
    18.50  &  107.0  &    2.9 \\
    19.01  &   83.6  &    2.3 \\
    19.48  &   79.7  &    2.2 \\
    20.00  &   67.2  &    1.9 \\
\hline
\end{tabular}
\end{table}
\section{Particle-$\gamma$ data on $^{90}$Y: Setup and method}
\begin{figure}[h]
\begin{center}
\includegraphics[width=1.\columnwidth]{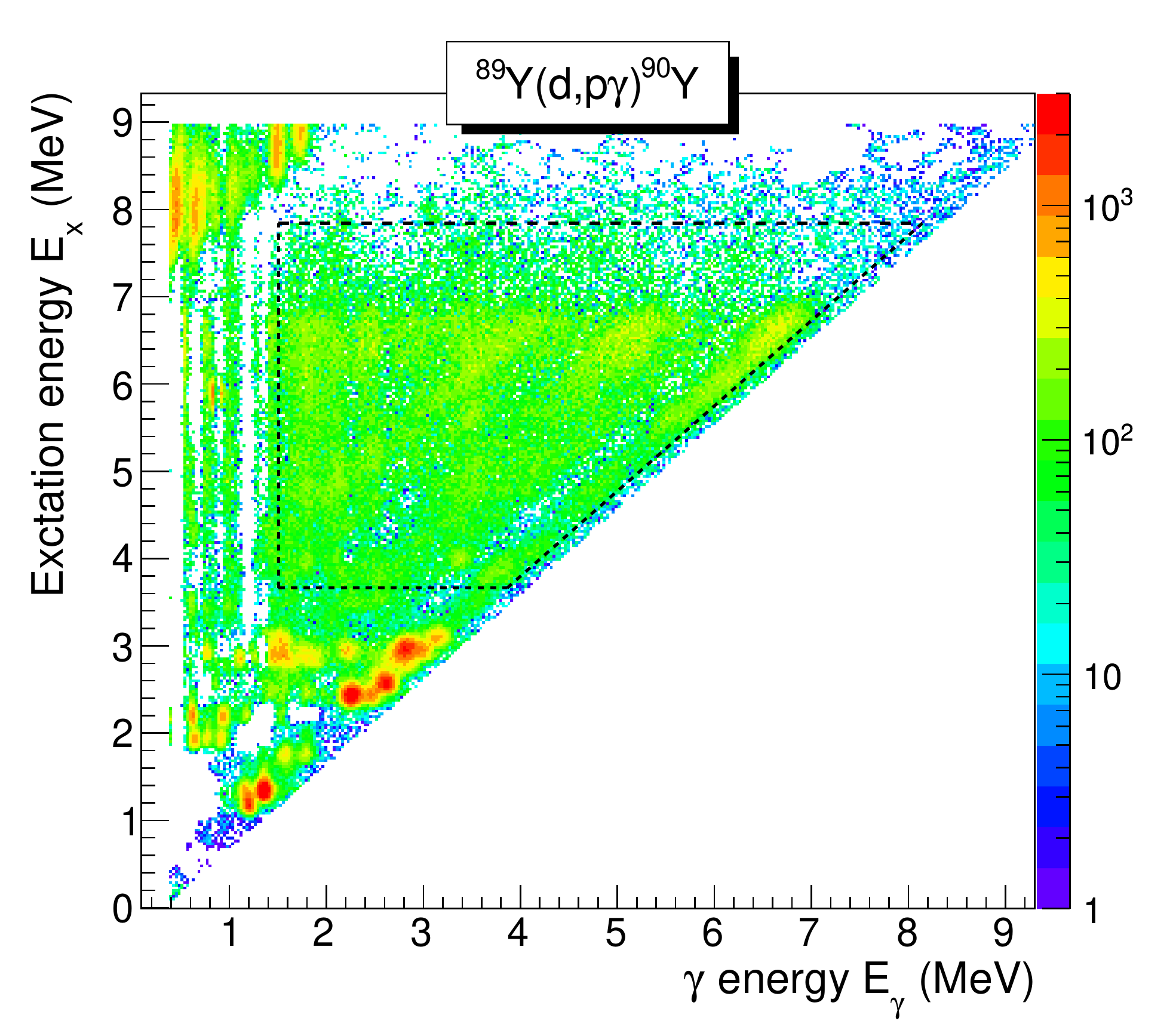}
\caption{(Color online) The primary $\gamma$-ray spectra as function of excitation energy, $E_x$, for the ($d,p\gamma$)$^{90}$Y data set. The dashed lines indicate the region used in the further analysis. \label{fig:primary}}
\end{center}
\end{figure}

The experiment probing the $\gamma$SF below $S_{\textrm{n}}$ was performed at the Oslo Cyclotron Laboratory (OCL), utilizing a deuteron beam of 13 MeV. The beam impinged on a natural $^{89}$Y target with thickness 2.25 mg/cm$^2$. Details about the experimental setup and analysis of the data are provided in Ref. \cite{guttormsen_yttrium_2014}.  Particle-$\gamma$ coincidences were measured with the particle-telescope system SiRi\cite{Guttormsen2011168} and the NaI(Tl) scintillator array CACTUS\cite{cactus} at OCL. The (d,p)-channel of the experiment was selected using $\Delta E-E$ technique. From the coincidence data, the primary $\gamma$-ray spectra, as shown in Fig.\ref{fig:primary}, for the excitation energies, $E_x$, was extracted using the iterative method described in Ref. \cite{schiller2000}. The primary spectra represent the distribution of the first emitted $\gamma$-rays in cascades from a given excitation energy range. The $\gamma$ transmission coefficient, $\mathfrak{T}(E_{\gamma})$, is assumed to depend only upon the energy of the emitted primary $\gamma$-ray, in keeping with the Brink hypothesis \cite{Brinkthesis,GuttormsenPRL2016}. In that case, the primary matrix can be factorized into two multiplicative functions as follows
\begin{equation} \label{eq:rhotausep}
P(E_{\gamma},E_x) \propto \rho(E_x - E_{\gamma}) \mathfrak{T}(E_{\gamma}),
\end{equation}
where $ \rho(E_x - E_{\gamma})$ is the nuclear level density at the excitation energy of the nucleus after a $\gamma$-ray with energy $E_{\gamma}$ has been emitted and $\mathfrak{T}(E_{\gamma})$ is the transmission coefficient. 

\subsection{Extraction of level density and $\mathfrak{T}(E_{\gamma})$}
\label{sec:exp}
From the distribution of primary $\gamma$-rays as function of excitation energy,
we get simultaneously information on both the nuclear level density (NLD) and 
$\gamma$-transmission coefficient~\cite{schiller2000}. 
The limits for extraction used in this work are: $E_\gamma^{min}$ = 1.51 MeV, $E_x^{min}$ = 3.67 MeV, and $E_x^{max}$ = 7.84 MeV. Although $E_x^{max}$ is higher than the neutron separation energy $S_n = 6.857$ MeV by approximately 1 MeV, we ensure that we are not using gamma spectra contaminated with gamma decay events from the ($d,pn\gamma$)$^{89}$Y channel by setting $E_\gamma^{min}$ = 1.51 MeV. The $E_x,E_\gamma$ matrix is previously shown in Ref.~\cite{guttormsen_yttrium_2014}. 

The obtained reduced $\chi^2$ is 2.8. Note that we do not attempt to
correct for Porter-Thomas fluctuations~\cite{porterthomas1956}, which are expected to be significant for nuclei with low level density. The excitation-energy resolution is $\approx 120$ keV (FWHM), determined from the width of the ground-state proton peak.

\begin{figure*}[!hbt]
\begin{center}
\includegraphics[clip,width=1.8\columnwidth]{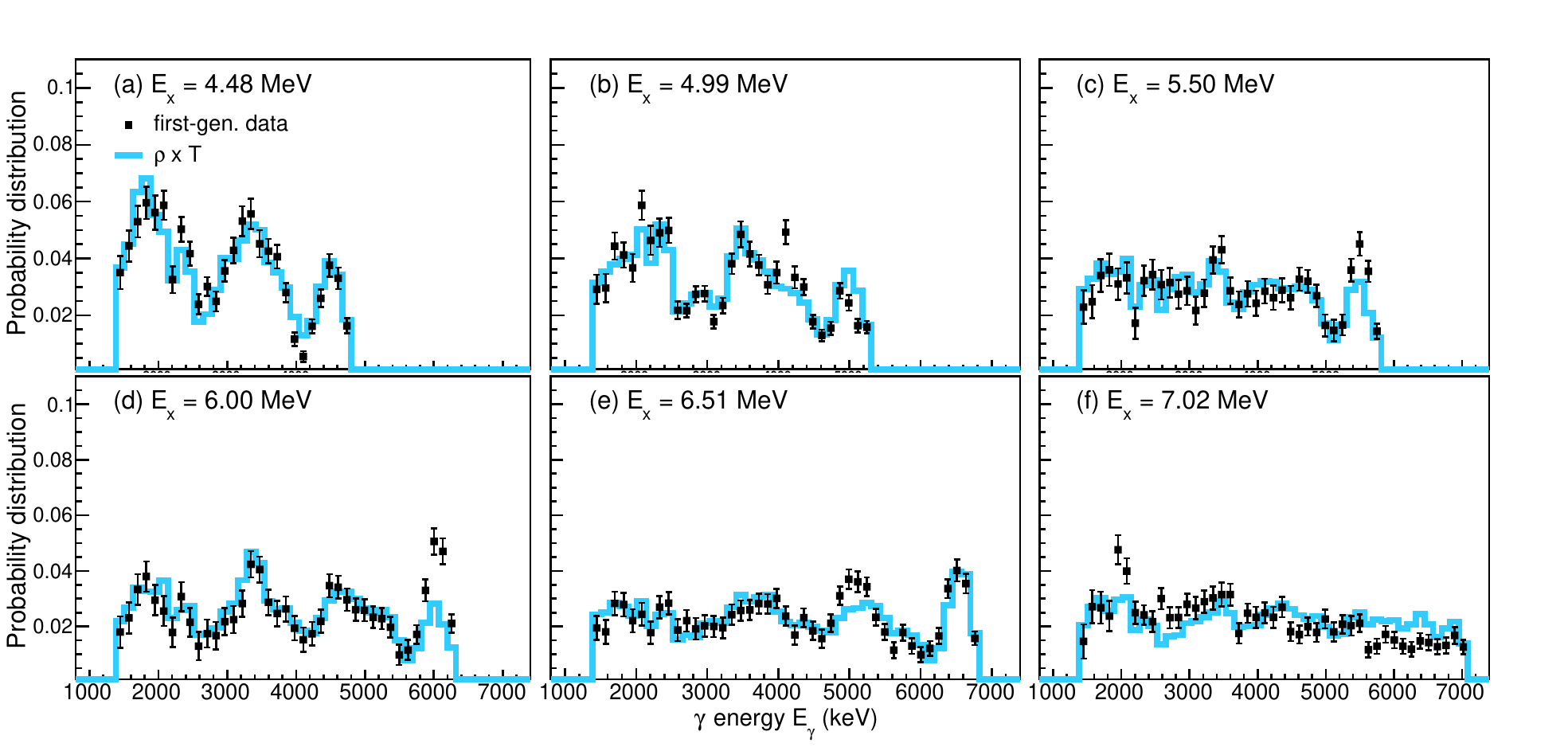}
%
\caption{(Color online). Data of primary $\gamma$ rays for several excitation-energy gates compared to the calculated result using the extracted $ \rho(E_x - E_{\gamma})$ and $\mathfrak{T}(E_{\gamma})$.}
\label{fig:doesitwork}
\end{center}
\end{figure*}
In Fig.~\ref{fig:doesitwork}, we test how well the functions $\rho(E_x-E_\gamma)$ and $\mathcal{T}(E_\gamma)$ extracted from the whole region within $E_\gamma^{min},E_x^{min},E_x^{max}$ reproduce \textit{individual} primary spectra from 127-keV $E_x$ bins.  In general, the product $\rho \times \mathcal{T}$ reproduces the data points well. Some data points do however deviate from the product by several orders of magnitude. This is likely to be due to Porter-Thomas fluctuations of transitions to individual or a few levels, as mentioned above. Note that the error bars in Fig.~\ref{fig:doesitwork} include statistical errors and systematic errors from the unfolding procedure and the extraction of primary $\gamma$ rays~\cite{schiller2000}. 

\subsection{Normalization of level density and $\gamma$-ray strength function}
\label{sec:norm}
As only the functional form is uniquely determined through the above mentioned fit procedure, the common slope and the absolute scales of the NLD and $\gamma$-transmission coefficient, respectively, are found by normalizing to auxiliary data. 

\subsubsection{Level density}
\label{sec:nld}
For the level density, we normalize to known, discrete levels~\cite{NNDC} at low excitation energy, where the level scheme is considered complete. In the case of $^{90}$Y, we normalize to the discrete levels (binned in 127-keV excitation-energy bins as our data points) for the range $E_x = 0.88 - 1.89$ MeV. 

Close to the neutron separation energy, $S_n$, we utilize neutron-resonance data for 
estimating the total level density $\rho(S_n)$ at that energy. For $^{90}$Y, we 
take the average $s$-wave resonance spacing $D_0$ from Ref.~\cite{mughabghab2006} of 4790(300) eV. To calculate $\rho(S_n)$ for all spins, not just the spins reached via $s$-wave 
neutron capture, we make use of the Hartree-Fock-Bogoliubov plus combinatorial 
(HFB+comb.) calculations of Goriely et al.~\cite{goriely2008} tuned to reproduce the $D_0$ value at $S_n$, using a shift $\delta$ and a slope correction $\alpha$ (see Eq.~(9) in Ref.~\cite{goriely2008}). 
We note that the spin distribution of these calculations are fully compatible with the average spin $\left< J \right>_{\mathrm{exp}} \approx 3.4$ at low $E_x \approx 1$ MeV for $^{90}$Y. 
We take the lower limit to be the highest value of $D_0$ (corresponding to the lowest level density) and vice versa, see Table~\ref{tab:par1}, and propagate the errors quadratically. 
Note that the previous normalization of $^{90}$Y in Ref.~\cite{guttormsen_yttrium_2014} is fully compatible with the lower limit of the present normalization.

\begin{table}[bt]
\begin{center}
\caption{NLD and $\gamma$SF normalization parameters for $^{90}$Y. The parameters $\alpha$
 and $\delta$ are used for matching the HFB+comb. calculations with the $D_0$ value as shown
 in Eq.~(9) of Ref.~\cite{goriely2008}.}
\begin{tabular}{lcccccc}
\hline
\hline
        & $D_0$		& $\rho(S_n)$ 	& $\rho_{\mathrm{red}}(S_n)$ & $\alpha$   &$\delta$ & $\left< \Gamma_{\gamma 0}\right>$ \\
        &  (eV)     & (MeV$^{-1}$)  & (MeV$^{-1}$)				 &(MeV$^{-1/2}$)& (MeV) & (meV)                             \\
\hline
Low 	& 5090      & 4924       	& 3689						 &-0.3763      & -0.269 & 101           \\
Middle  & 4790      & 5232         	& 3920						 &-0.3527      & -0.269 & 168 \\
High 	& 4490      & 5582         	& 4182 						 &-0.3275      & -0.269 & 302 \\
\hline
\hline
\end{tabular}
\label{tab:par1}
\end{center}
\end{table}
\begin{figure}[bt]
\begin{center}
\includegraphics[clip,width=1.\columnwidth]{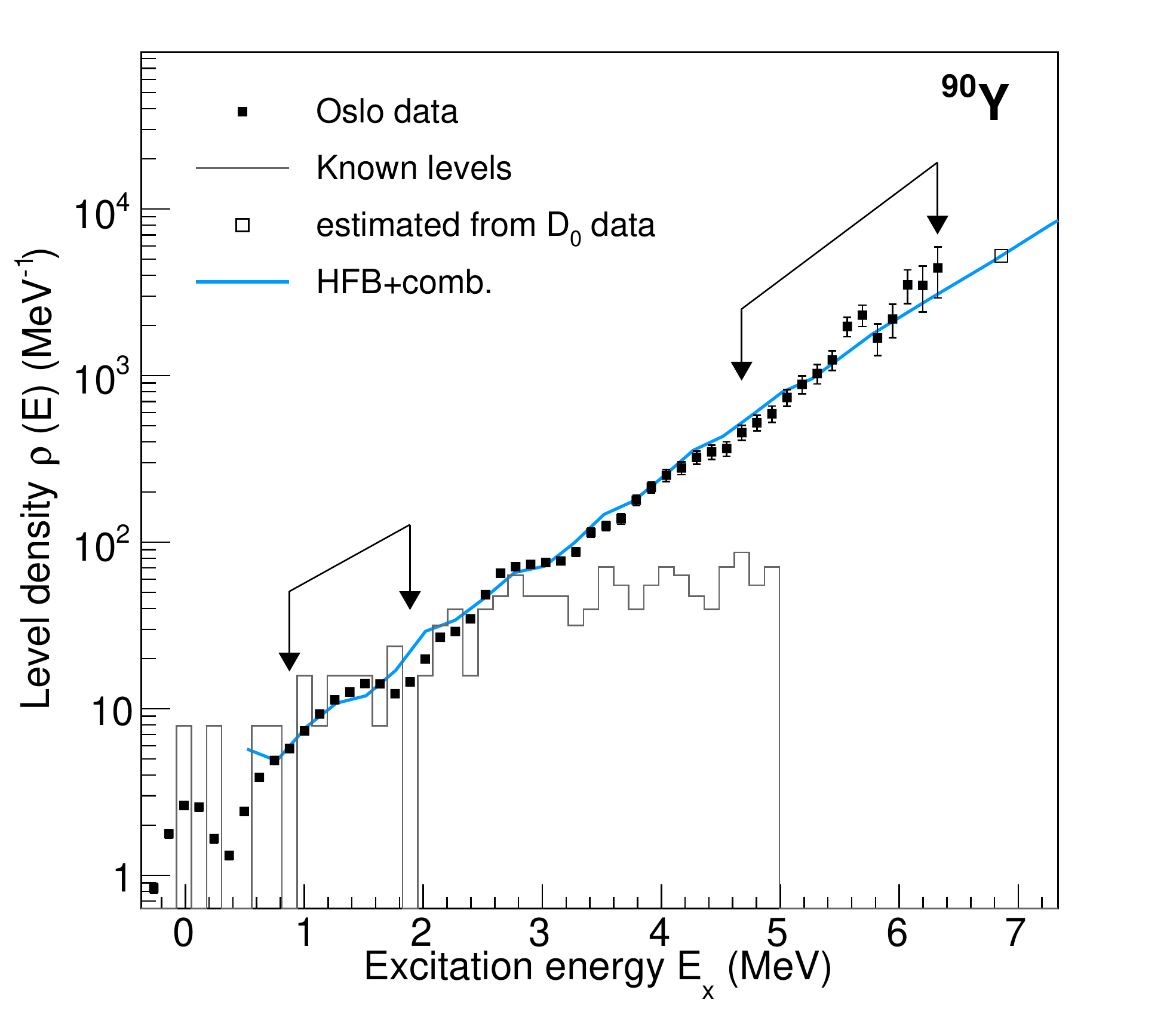}
\caption{(Color online) Normalized level density of $^{90}$Y (see text).
The data points within the arrows are used for normalization. Error bars include statistical errors, systematic errors from the unfolding and extraction of the primary $\gamma$-ray spectra, and systematic errors from the normalization to the $D_0$ value.}
\label{fig:leveldensity90Y}
\end{center}
\end{figure}

Because the $^{89}$Y($d,p$) reaction does not populate high spins, and the slope of the NLD
is intertwined with the slope of the $\gamma$-transmission coefficient, we also estimate
a reduced NLD corresponding to a spin range representative of the one populated in the
experiment. 
From levels populated in previous ($d,p$) experiments~\cite{NNDC}, in particular levels given in table 3 of Ref.~\cite{michaelsenNPA} and 
levels identified in the present experiment, we estimate the spin range of the directly populated levels to be $J\approx 0-6$. 
Further, we take into account that our NLD is determined \textit{after} emission of one dipole transition carrying $L=1$, so that the spin range of the final levels is $J\approx 0-7$.
Using the spin distribution of the HFB+comb. calculations at $S_n$, the NLD for the spin range $J= 0-7$ corresponds to $\approx 75$\% of the total NLD for all spins at $S_n$. 
This reduced NLD, $\rho_{\mathrm{red}}(S_n)$,  (see Table~\ref{tab:par1}) gives us the slope for the $\gamma$-transmission coefficient. 
The normalized level density is shown in Fig.~\ref{fig:leveldensity90Y}.
This minor reduction of the slope of the $\gamma$-transmission coefficient is not crucial for the further analysis. In fact, the  $\gamma$-transmission coefficient obtained by assuming a full coverage of all spins available in the HFB+comb. calculations is well within the final systematic errors.

\subsubsection{$\gamma$-ray strength function}
\label{sec:gsf}
The slope of the $\gamma$-transmission coefficient is determined 
by normalizing the NLD to the reduced $\rho_{\mathrm{red}}(S_n)$ as described in the previous section. The absolute scale was found by use of the total, average radiative width 
$\left< \Gamma_{\gamma0} \right>$ as described in Ref.~\cite{voinov2001}:
Ref.~\cite{mughabghab2006} gives for $^{90}$Y a value $\left< \Gamma_{\gamma0} \right> = 134$ meV, without any estimate of the uncertainty. By closer inspection of the individual $\Gamma_{\gamma0}$ values listed, it is clear that a rather wide range of possible $\left< \Gamma_{\gamma0} \right>$ can be estimated. The values from Ref.~\cite{mughabghab2006} are provided in Table~\ref{tab:gamma}. 
\begin{table}[ht]
\begin{center}
\caption{Individual $\Gamma_{\gamma0}$ widths for $^{90}$Y as listed in Ref.~\cite{mughabghab2006}.
As $^{89}$Y as $J^\pi = 1/2^-$ in the ground state, $s$-wave capture gives $J=0^-,1^-$ levels in $^{90}$Y.}
\begin{tabular}{rcr}
\hline
\hline
$E_n$       & $J$	& $\Gamma_{\gamma0}$  	\\
(keV)       &       & (meV)				  	\\
\hline
$-0.251$	& 1     & 126    				\\   	             
$2.598$     & 1     & 131(10)         		\\
$7.498$ 	& 0     & 116(12)         		\\
$11.59$ 	& 0     & 542(64)         		\\
$13.78$ 	& 1     & 109(11)         		\\
$15.23$ 	& 0     & 92(9)         		\\
$26.40$ 	& 0     & 128(15)         		\\
$26.94$ 	& [1]   & 106(10)         		\\
$29.65$ 	& 1     & 151(15)         		\\
$38.06$ 	& 1     & 174(20)         		\\
\hline
\hline
\end{tabular}
\label{tab:gamma}
\end{center}
\end{table}

Making an average of all the values in Table~\ref{tab:gamma} we get 
$\left< \Gamma_{\gamma0} \right> = 168$ meV. Calculating the unbiased standard deviation, 
this yields 134 meV. However, by removing the abnormal 11.59-keV resonance with $\Gamma_{\gamma0} = 542(64)$ meV from the average, we obtain $\left< \Gamma_{\gamma0} \right> = 126(25)$ meV. 
Based on these considerations we estimate $\left< \Gamma_{\gamma0} \right> = 168_{-67}^{+134}$ meV, so that the lower(upper) limit is given by the results excluding(including) the resonance with the largest width (see also Table~\ref{tab:gamma}). 

\begin{figure}[!t]
 \begin{center}
 \includegraphics[clip,width=1.\columnwidth]{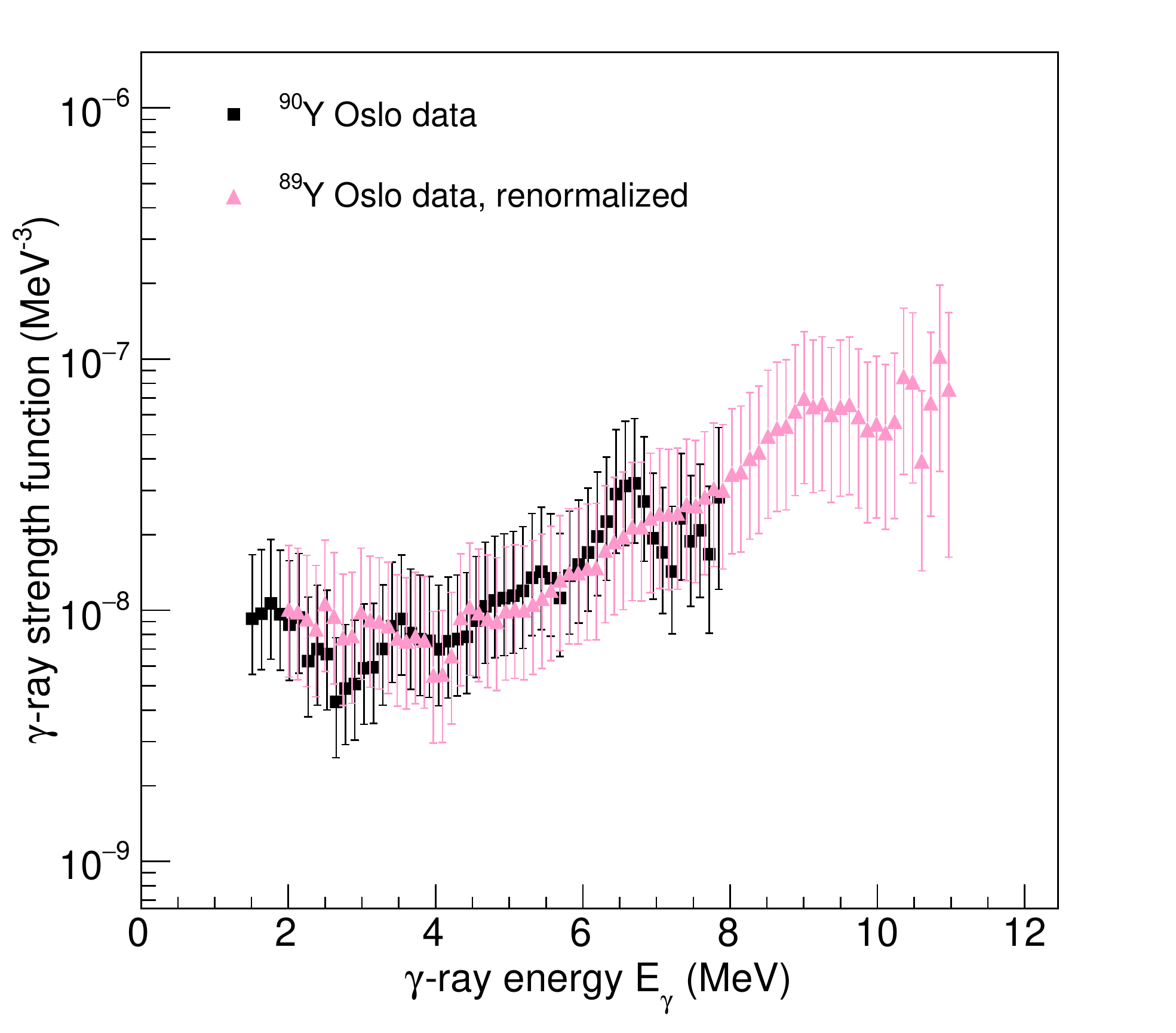}
 \caption{(Color online) Normalized $\gamma$SF of $^{90}$Y 
 shown together with the renormalized $^{89}$Y data from Ref.~\cite{PhysRevC.93.045810}. The error bars include statistical errors,  systematic uncertainties from the unfolding and extraction of primary $\gamma$-ray spectra, and systematic uncertainties from the normalization.}
 \label{fig:gsfdata}
 \end{center}
 \end{figure}

From the normalized transmission coefficient, the $\gamma$SF is determined by
\begin{equation}
f(E_\gamma) = \frac{\mathcal{T}(E_{\gamma})}{2\pi E_\gamma^3},
\end{equation}
since dipole radiation dominates in the considered $E_x$ region~\cite{kopecky2017,larsen2013}. 
The normalized $\gamma$SF is shown in Fig.~\ref{fig:gsfdata}. 

\subsection{Re-evaluation of the $^{89}$Y level density and $\gamma$-ray strength function}
\label{sec:gsf89Y}
For completeness, we have also re-evaluated the normalization of the $^{89}$Y data from Ref.~\cite{PhysRevC.93.045810}. 
As for $^{89}$Y, 
the HFB+comb calculations of Ref.~\cite{goriely2008} reproduce well
the average spin at low excitation energies, $\left< J \right>_{\mathrm{exp}} \approx 3.3$ around $E_x \approx 2.2$ MeV. 
For the re-normalization, we use the HFB+comb. calculations with the following parameters keeping the shift $\delta = 0.$ MeV in all cases: middle normalization with $D_0 = 121$ eV at $S_n = 11. 482$ MeV, $\alpha = 0.0$ MeV$^{-1/2}$; high normalization with $D_0 = 100$ eV, $\alpha = 0.0551$ MeV$^{-1/2}$ ; low normalization with $D_0 = 143$ eV, $\alpha = -0.0505$ MeV$^{-1/2}$.

\begin{figure}[!t]
 \begin{center}
 \includegraphics[clip,width=1.\columnwidth]{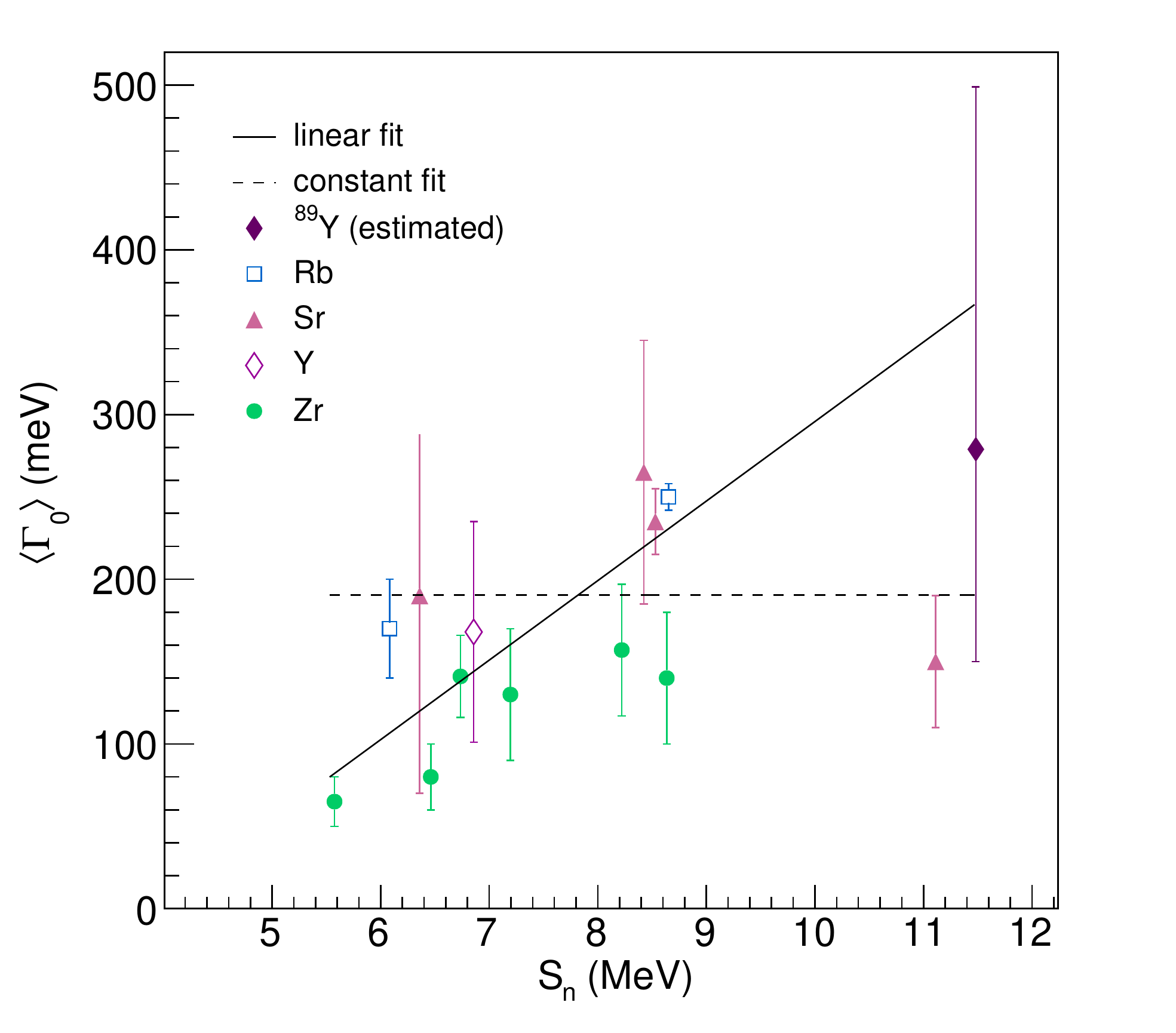}
 \caption{(Color online) Fit of available 
 $\left< \Gamma_{\gamma 0} \right>$ data for Rb, Sr, Y and Zr isotopes taken from Refs.~\cite{mughabghab2006,guttormsen2017}.}
 \label{fig:Ggfit}
 \end{center}
 \end{figure}

For the normalization of the $\gamma$SF of $^{89}$Y, we have considered all available 
$\left< \Gamma_{\gamma 0} \right>$ data for Rb, Sr, Y and Zr isotopes from Ref.~\cite{mughabghab2006}. 
For $^{91,92}$Zr, we use the adopted values from Ref.~\cite{guttormsen2017}.
As noted for $^{90}$Y, the $\Gamma_{\gamma 0}$ values for individual $s$-wave resonances vary considerably, as do the estimated averages. 
With the aim of catching the spread in the 
$\left< \Gamma_{\gamma 0} \right>$ data, we have fitted simple polynomials to the available data as shown in Fig.~\ref{fig:Ggfit}. 
From these fits, we estimate $\left< \Gamma_{\gamma 0} \right> = 279_{-129}^{+220}$ meV for $^{89}$Y, where the central value is taken as the average of the linear and constant fit, the lower limit corresponds to the one estimated in Ref.~\cite{PhysRevC.93.045810}, and the upper limit is taken as 79\% above the central value (as estimated for $^{90}$Y). 
The present central value is considerably higher and with larger errors than the previous value from Ref.~\cite{PhysRevC.93.045810} of 150(38) meV. 
The resulting renormalized $\gamma$SF of $^{89}$Y is shown in Fig.~\ref{fig:gsfdata}.

\section{Comparison of data}
\label{sec:data}
\begin{figure}
\includegraphics[width=0.48\textwidth]{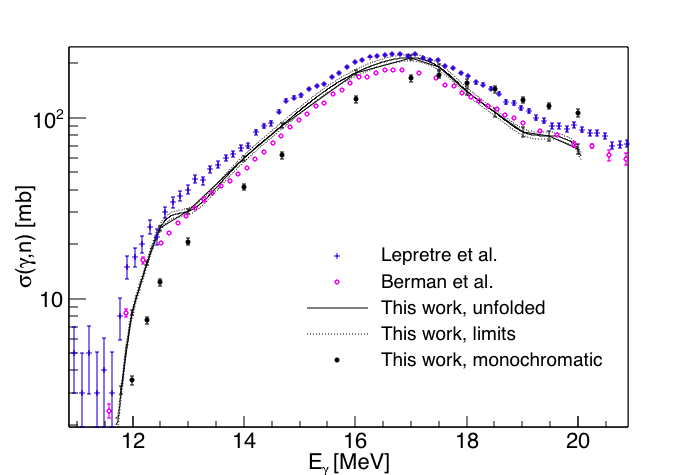}
\caption{The variation in the cross section under the monochromatic assumption was determined by the simulation procedure described in the text (for $10^7$ samples). The upper and lower limits on the unfolded cross section corresponds to unfolding the monochromatic cross section $\pm 1 \textrm{SE}$, where $\textrm{SE}$ is the standard error found in \ref{sec:errorprop}. For comparison, previous experimental results by Berman et al. \cite{berman1967} and Lepretre et al. \cite{lepretre1971} are also displayed. \label{fig:gnfinal}}
\end{figure}

\begin{figure}[!t]
 \begin{center}
 \includegraphics[clip,width=1.\columnwidth]{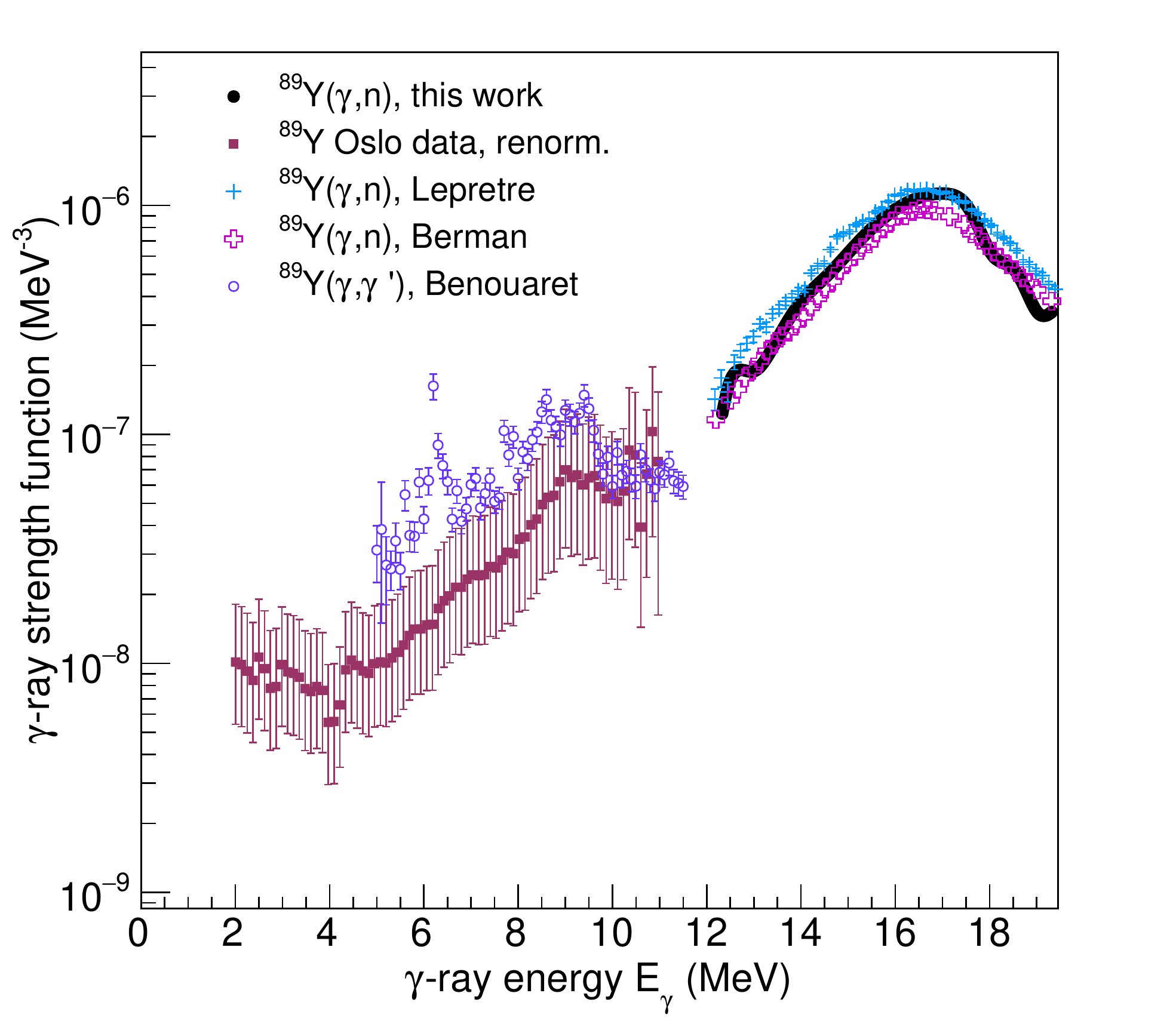}
 \caption{(Color online) Normalized $\gamma$SF of $^{89}$Y 
 shown together with new $^{89}$Y($\gamma,n$) data from this work, and $^{89}$Y($\gamma,n$) data from 
 Refs.~\cite{berman1967,lepretre1971} as well as $^{89}$Y($\gamma,\gamma^\prime$) data from Ref.~\cite{benouaret2009}.}
 \label{fig:allgsfdata}
 \end{center}
 \end{figure}

Our results for the $^{89}$Y($\gamma$,n) cross section are here compared to existing data for this reaction in Fig. \ref{fig:gnfinal}. The measurements of Berman et al. \cite{berman1967} and Lepretre et al. \cite{lepretre1971} show a significant discrepancy for the whole energy range probed by the two experimental campaigns. Our measured cross section represents an intermediate value, but somewhat closer in value to Lepretre et al's result for $E_{\gamma}$ less than approximately 18 MeV. For $E_{\gamma}>$ 18 MeV, our results are compatible with the results of Lepretre et al. 
To compare all available data for $^{89}$Y, the $^{89}$Y($\gamma,n$) cross section data from Refs.~\cite{berman1967,lepretre1971} and the new $^{89}$Y($\gamma,n$) data from this work are converted into $\gamma$SF using the principle of detailed balance~\cite{blatt1952} by the relation~\cite{RIPL}
\begin{equation}
f(E_\gamma) = \frac{\sigma_{(\gamma,n)}(E_\gamma)}{3\pi^2 \hbar^2 c^2 E_\gamma},
\label{eq:crossGSF}
\end{equation}
again assuming that dipole radiation is dominant. 
These data are also shown in Fig.~\ref{fig:allgsfdata}. 
We note that ($\gamma,n$) cross section data are not a good measure for the $\gamma$SF close to neutron threshold due to threshold effects of the neutron emission (see, \textit{e.g}, Ref.~\cite{utsunomiya2009}); most importantly the competition of the neutron channel with the $\gamma$ channels. Hence, the ($\gamma,n$) data closest to $S_n$ are not used in the following. 
Eq.~(\ref{eq:crossGSF}) is also used for transforming $^{89}$Y($\gamma,\gamma^\prime$) cross sections from Ref.~\cite{benouaret2009} into $\gamma$SF. All the data are shown together in Fig.~\ref{fig:allgsfdata}.  

In general, the $^{89,90}$Y $\gamma$SFs increase as a function of $\gamma$-ray energy. 
This is expected as we are measuring the low-energy tail of the giant dipole resonance
(GDR)~\cite{dietrich1988}, which in this case is centered around $E_\gamma \approx 17$ MeV, 
and is represented by the photoneutron data. 
In the $^{90}$Y Oslo data, we note that there is a peak at $E_\gamma\approx 6.6$ MeV. 
This is likely due to strong $M1$ spin-flip transitions for the neutron configuration $\nu(0g_{9/2}^{-1} 0g_{7/2}^1)$. 
Such spin-flip transitions have been measured recently 
in a photon-scattering experiment on the $N=50$ isotone $^{90}$Zr at the HI$\gamma$S facility~\cite{rus13}. Also, in a previous measurement of the ($n,\gamma$)$^{90}$Y reaction by Raman \textit{et al.}~\cite{raman1981}, it was found that the 2.6-keV resonance decays via (a) strong $M1$ transition(s). 

Moreover, we observe an increase at decreasing $\gamma$ energies for $E_\gamma \lesssim 3$ MeV for both $^{89,90}$Y. This feature has been seen in many nuclei since the first observation in the iron isotopes~\cite{voinov2004}, where it was recently shown to also be dominated by dipole transitions~\cite{larsen2013,simon2016,larsen2017}. 

The physical mechanism causing the low-energy enhancement is, however, still unclear, despite its presence in many nuclei, with the deformed $^{151,153}$Sm being the heaviest cases so far~\cite{simon2016}.
Within the thermal-continuum quasiparticle random phase approximation (TCQRPA), the low-energy enhancement is explained as being due to $E1$ transitions~\cite{litvinova2013}, while shell-model calculations predict an increase in strength for low-energy $M1$ transitions~\cite{schwengner2013,brown2014,schwengner2017}, even when the $E1$ component is calculated as well~\cite{sieja2017}. 
A recent Compton polarization measurement by Jones \textit{et al.} using the GRETINA array~\cite{Jones2018} shows a slight bias towards $M1$ transitions. However, the data statistics do not allow to draw significant conclusions about the source of the low-energy enhancement in $^{56}$Fe. Admixtures of $E1$ and  $M1$ transitions cannot be ruled out. 

In the following section (Sec. \ref{sec:shell}) we discuss predictions of the $\gamma$SF from shell-model calculations on $^{90}$Y for the quasi-continuum region, as well as recent calculations within the quasi-particle random phase approximation (QRPA) for the $E1$ and $M1$ strength built on the ground state. In addition, we  compare with fits using phenomenological models. 
\section{Model descriptions of the $\gamma$SF}
\label{sec:shell}
\subsection{Shell-model calculations}
 \begin{figure*}[!ht]
 \begin{center}
 \includegraphics[clip,width=2.\columnwidth]{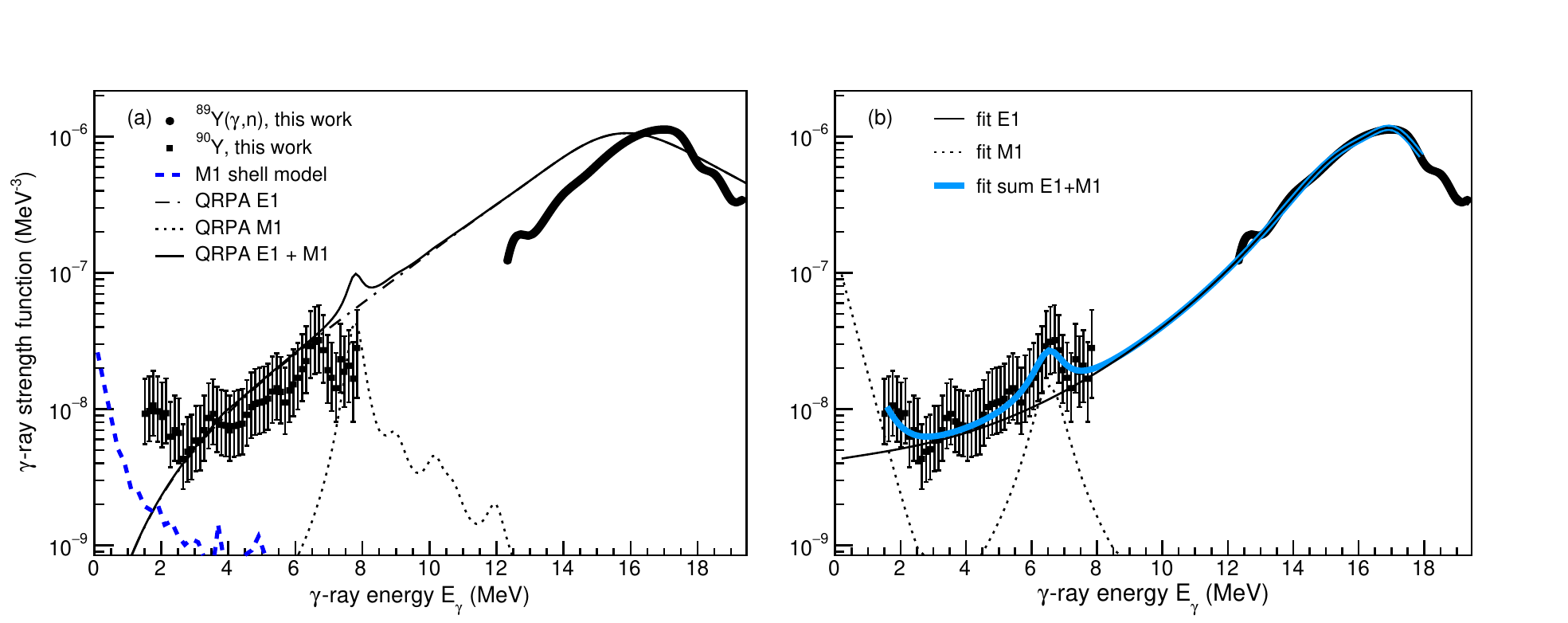}
 \caption{(Color online) Comparison of data and microscopic calculations of the
 dipole $\gamma$SF (a) and phenomenological models (b).}
 \label{fig:strength_calc}
 \end{center}
 \end{figure*}
\begin{table*}[ht]
\begin{center}
\caption{Parameters found from the model fits of $f_{\mathrm{tot}}^1$ to the $\gamma$SF of $^{90}$Y and ($\gamma,n$) data.}
\begin{tabular}{lcclcllclllcc}
\hline
\hline
Norm. 	& $E_{E1,1}$ & $\Gamma_{E1,1}$& $\sigma_{E1,1}$  & $E_{E1,2}$ & $\Gamma_{E1,2}$ & $\sigma_{E1,2}$ & $T_f$    &$E_{M1}$ & $\Gamma_{M1}$ & $\sigma_{M1}$ & $C$           & $\eta$	     \\
        & (MeV)      & (MeV)          &  (mb)            & (MeV)      & (MeV)           &  (mb)           & (MeV)    & (MeV)   & (MeV)         &(mb)           & 10$^{-7}$(MeV$^{-3}$) &(MeV$^{-1}$) \\
\hline
Low      & 16.1(1)   & 3.85(5)        & 115(7)           & 17.1(1)    & 1.99(10)        & 153(7)          & 0.73(2)  & 6.54(3) & 0.56(10)      & 0.75(11)      & 0.9(2)        & 2.1(1) \\
Middle   & 16.2(1)   & 3.71(6)        & 131(8)           & 17.1(1)    & 1.78(10)        & 149(9)          & 0.96(2)  & 6.56(4) & 1.03(19)      & 1.14(16)      & 1.5(3)        & 2.1(1) \\
High     & 16.1(1)   & 3.00(5)        & 146(10)          & 17.1(1)    & 1.60(8)         & 180(10)         & 1.45(3)  & 6.60(4) & 1.34(16)      & 2.16(20)      & 2.8(5)        & 2.1(1) \\
\hline
\hline
\end{tabular}
\label{tab:gsfpar}
\end{center}
\end{table*}

The shell-model calculations were performed with the RITSSCHIL code \cite{zwa85} with a model space consisting of the $\pi(0f_{5/2}, 1p_{3/2}, 1p_{1/2}, 0g_{9/2})$ proton orbits and the
$\nu(0g_{9/2}, 1d_{5/2}, 0g_{7/2})$ neutron orbits relative to a $^{68}$Ni core. The same configuration space was also applied in our earlier study of the $M1$ strength functions in $^{94,95,96}$Mo and $^{90}$Zr \cite{schwengner2013}. In the present calculations, two protons were allowed to be lifted from the $(fp)$ shell to the $0g_{9/2}$ orbit and two neutrons from the $0g_{9/2}$ to the $1d_{5/2}$ orbit. This resulted in dimensions up to 29000. The exclusion of an
occupation of the $\nu(0g_{7/2})$ orbit suppresses the spin-flip peak formed mainly by $1^+ \rightarrow 0^+$ transitions with energies, $E_{\gamma}$, around 7 MeV \cite{schwengner2013,rus13}, but turned out to have no significant influence on the low-energy part of the strength function.
 
The calculations included states with spins from $J$ = 0 to 10 for $^{90}$Y. 
For each spin the lowest 40 states were calculated. Reduced transition probabilities $B(M1)$ were
calculated for all transitions from initial to final states with energies $E_f < E_i$ and spins following the usual dipole selection rules. For the minimum and maximum $J_i$, the cases $J_f = J_i - 1$ and $J_f = J_i + 1$, respectively, were excluded. This resulted in more than 32000 $M1$ transitions for each parity $\pi = +$ and $\pi = -$, which were sorted into 100 keV bins according to their transition energy $E_\gamma = E_i - E_f$. The average $B(M1)$ value for one energy bin was obtained as the sum of all $B(M1)$ values divided by the number of transitions within this bin.

The $M1$ strength functions were deduced using the relation 
\begin{equation}
f_{M1}(E_i,E_\gamma, J, \pi)= 
a\left<B(M1,E_i,E_\gamma, J, \pi)\right>\cdot\rho(E_i, J, \pi).
\end{equation}
This corresponds to the relation given in Ref.~\cite{bartholomew1972}
using $B(M1) = a \Gamma E^{-3}$ where $a=16\pi/9 (\hbar c)^{-3}$. They were calculated by multiplying the ${B(M1)}$ value in $\mu^2_N$ of each
transition with $11.5473 \times 10^{-9}$ times the level density at the energy, as determined by these calculations,
of the initial state $\rho(E_i)$ in MeV$^{-1}$ and deducing averages in energy
bins as done for the $\left<B(M1)\right>$ values (see above) and averaging over $J$, $\pi$ and $E_i$. 
When calculating the strength functions, gates were set on the excitation
energy $E_i$ that correspond to the ones applied in the analysis of the
experimental data, namely 3.67 - 7.84 MeV (see Sec.~\ref{sec:exp}).
The resulting $M1$ strength function for $^{90}$Y is shown in Fig.~\ref{fig:strength_calc}. The low-energy
behavior shows an increase at low energies similar to that of the strength functions calculated for the neighboring nuclei $^{94,95,96}$Mo, $^{90}$Zr \cite{schwengner2013} and for $^{56,57}$Fe \cite{brown2014}.

The low-energy enhancement of $M1$ strength is in the shell model picture caused by transitions between the several close-lying states of all considered spins located above the yrast line in
the transitional region to the quasi-continuum of nuclear states.  Inspecting the wave functions, one finds large $B(M1)$ values for transitions between states that contain a large component (up to about 50\%) of the same configuration with broken pairs of both protons and neutrons in high-$j$ orbits, whereas states containing only proton excitations or only neutron excitations are not depopulated by strong $M1$ transitions. 

The largest $M1$ matrix elements connect configurations with the spins of high-$j$ protons
re-coupled with respect to those of high-$j$ neutrons to the total spin $J_f = J_i, J_i \pm 1$. The corresponding main configurations for negative-parity states in $^{90}$Y are 
$\pi(1p_{1/2}^1) \nu(0g_{9/2}^{-1} 1d_{5/2}^2)$ or 
$\pi(1p_{1/2}^1) \nu(0g_{9/2}^{-2} 1d_{5/2}^3)$ and by additional proton
excitations within the $(fp)$ shell, i.e.
$\pi[(0f_{5/2}, 1p_{3/2})^{-1} 1p_{1/2}^2] \nu(0g_{9/2}^{-1} 1d_{5/2}^2)$
and also proton excitations over the subshell gap at $Z$ = 40,
$\pi[(0f_{5/2}, 1p_{3/2})^{-1} 1p_{1/2}^0 0g_{9/2}^2]\nu(0g_{9/2}^{-1} 1d_{5/2}^2)$.
The positive-parity states require the excitation of an $(fp)$ proton to the
$0g_{9/2}$ orbit, for example 
$\pi(1p_{3/2}^{-1} 1p_{1/2}^1 0g_{9/2}^1) \nu(0g_{9/2}^{-1} 1d_{5/2}^2)$.
The orbits in these configurations have large $g$ factors with opposite signs
for protons and neutrons. Combined with specific relative phases of the proton
and neutron partitions they cause large total magnetic moments. 

\subsection{QRPA calculations}
As $E1$ transitions are out of reach within the framework of the shell model in this case,
we have employed recent QRPA calculations based on the D1M Gogny force taken from 
Ref.~\cite{martini2016}. These calculations give the $E1$ strength for
one-particle-one-hole excitations built on the ground state only, and are not
necessarily representative of the $E1$ strength in quasi-continuum. 
On the other hand, if the Brink hypothesis~\cite{brink1955} is approximately correct, the obtained $E1$ strength should be a good substitute for the quasi-continuum strength.
Further, also the ground-state $M1$ strength is obtained within the same framework~\cite{goriely2016}. The microscopic calculations including the shell-model results are shown together with  the data in Fig.~\ref{fig:strength_calc}. 

It is apparent that the QRPA $E1$ strength describes rather well the lower limit of the ($d,p\gamma$)$^{90}$Y data, while the GDR centroid is shifted towards lower $E_\gamma$ with respect to the ($\gamma,n$) data.  The QRPA $M1$ strength shows quite a bit of structure with a strong peak around $E_\gamma \approx 7.8$ MeV, consistent with the expected spin-flip transitions, and probably related to the peak seen in the $^{90}$Y data about 1 MeV lower in $E_\gamma$. 

As expected, the QRPA $M1$ strength shows no low-energy increase in strength as these are built up of ground-state excitations. In contrast, the shell-model calculations show a prominent
low-energy increase, although lower in absolute value than the ($d,p\gamma$)$^{90}$Y data. This indicates that the upbend in $^{90}$Y can be understood as relating to transitions between excited states in the quasi-continuum. 

\subsection{Phenomenological models}

We have used the phenomenological Generalized Lorentzian (GLo) model~\cite{kopecky_uhl_1990}, with a constant temperature of the final states $T_f$ in 
agreement with the Brink hypothesis~\cite{brink1955}. The GLo model is given by
\begin{align}
& f_{\rm GLo}^{E1}(E_{\gamma},T_f) = \frac{1}{3\pi^2\hbar^2c^2}\sigma_{E1}\Gamma_{E1} \times \\ \nonumber
& \left[\frac{ E_{\gamma} \Gamma(E_{\gamma},T_f)}{(E_\gamma^2-E_{E1}^2)^2 + E_{\gamma}^2 
	\Gamma (E_{\gamma},T_f)^2} + \;0.7\frac{\Gamma(E_{\gamma}=0,T_f)}{E_{E1}^3} \right],
\label{eq:GLO}
\end{align} 
with
\begin{equation}
\Gamma(E_{\gamma},T_f) = \frac{\Gamma_{E1}}{E_{E1}^2} (E_{\gamma}^2 + 4\pi^2 T_f^2).
\end{equation}
Here, the parameters $\Gamma_{E1}$, $E_{E1}$ and $\sigma_{E1}$ correspond to the width, 
centroid energy, and peak cross section of the GDR respectively. 

To simultaneously fit the $^{89}$Y($\gamma,n$) data from this work together with the 
($d,p\gamma$)$^{90}$Y data, we have used two GLo functions for the $E1$ strength
with a common temperature $T_f$ together with a Standard Lorentzian (SLo) function for the $M1$ spin-flip resonance, and an exponential function of the form $f_{\mathrm{upbend}}^{M1} = C\exp{-\eta E_{\gamma}}$. 
Although $^{89}$Y is considered to be a spherical nucleus, where only one GLo function would be assumed to be sufficient to describe the GDR, we find that our ($\gamma,n$) data display significant structures, and that the peak around $E_\gamma \approx 16-17$ MeV is rather flat. 
Hence, we introduce two GLo components to better reproduce the ($\gamma,n$) data.
We obtain the total dipole-strength fit function
\begin{equation}
f_{\mathrm{tot}}^1 = f_{\rm GLo1}^{E1} + f_{\rm GLo2}^{E1} + f_{\rm SLo}^{M1} + f_{\mathrm{upbend}}^{M1},
\label{eq:glofit}
\end{equation}
with, in principle, 12 free parameters in the fit. 

To restrict the temperature parameter, we first 
performed an individual fit of the two GLo components to the present $^{89}$Y($\gamma,n$) data in the range of $E_\gamma = 14.0-18.0$ MeV and the $^{89}$Y($d,p\gamma$)$^{90}$Y data of this work in the range of $E_\gamma = 1.5-7.9$ MeV. From this fit of the $E1$ component, we determine $T_f$ and fix it in the next fit where we include the $f_{\rm SLo}^{E1}$ and $f_{\mathrm{upbend}}^{M1}$ terms, so that there are in practice 11 free parameters. 
We performed three different fits for the lower, middle and upper normalizations.  
The obtained parameters from the three fits to the upper, lower and middle $f_{\mathrm{tot}}^{1}$ are listed in Table~\ref{tab:gsfpar}.

We find that the centroids $E_{E1,1}, E_{E1,2}$ are similar regardless of which normalization is used for the $^{90}$Y data, as expected since these centroids are mainly determined by the ($\gamma,n$) data. 
The other GLo parameters vary significantly from the fits to the low, middle and high normalization to compensate for the change in absolute value of the $^{90}$Y data. Further, the $M1$ spin-flip centroid is not sensitive to our choice of normalization, while the width and peak cross section vary according to the low, middle and high normalizations. 
The parameters for the exponential fit to the upbend indicate a stable slope of $\eta = 2.1(1)$ MeV$^{-1}$, while the constant $C$ again show a large spread in accordance with the normalization uncertainties. It is interesting that the $\eta$ parameter is similar to that found for $^{89}$Y: $\eta(^{89}\mathrm{Y}) \approx 2.5$ MeV$^{-1}$~\cite{PhysRevC.93.045810}. 
 
\section{Radiative neutron capture cross section and reaction rate}
\label{sec:talys}
 \begin{figure*}[!htb]
 \begin{center}
 \includegraphics[clip,width=1.9\columnwidth]{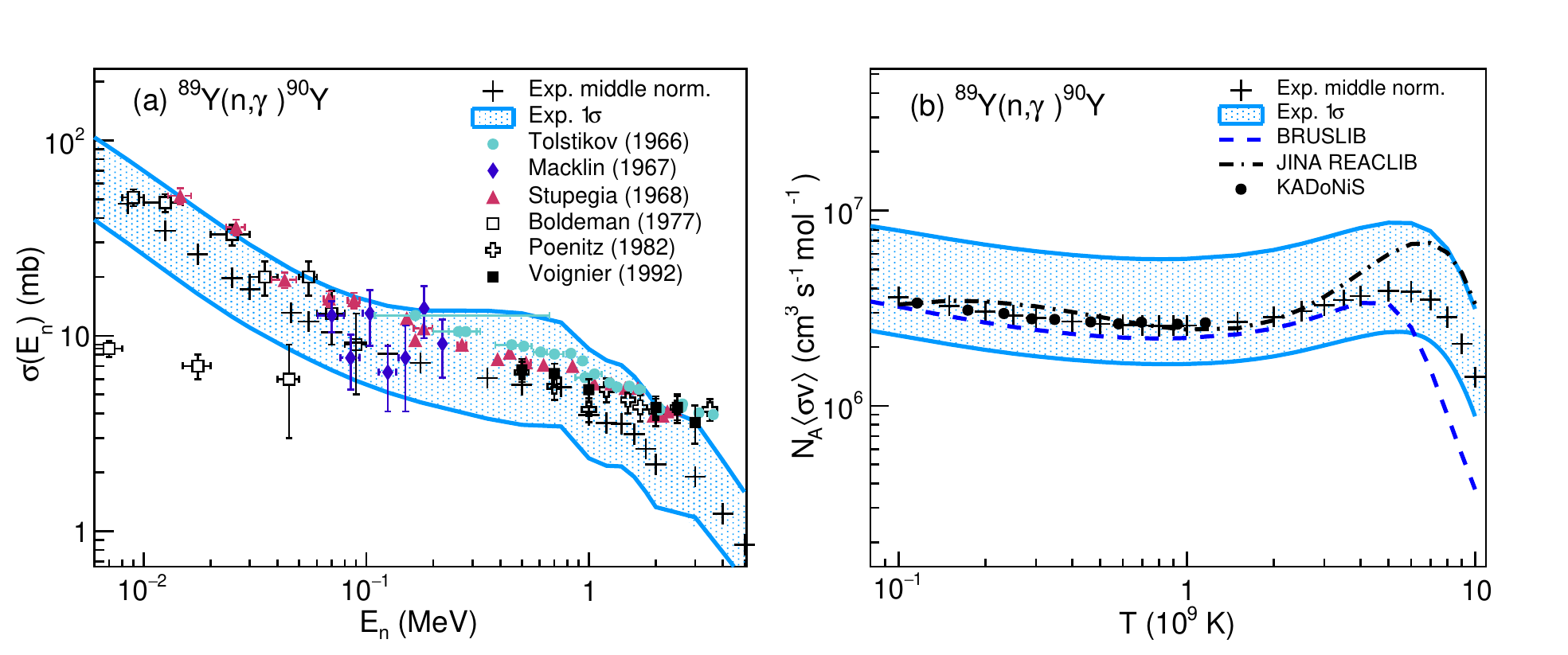}
 \caption{(Color online) Calculated $^{89}$Y($n,\gamma$)$^{90}$Y cross sections (a) compared
 to data from Refs.~\cite{tolstikov1966,macklin1967,stupegia1968,boldeman1977,poenitz1982,voignier1992}, 
 and the corresponding astrophysical reaction rates (b) compared with BRUSLIB~\cite{BRUSLIB} and 
 JINA REACLIB (kd02-v06)~\cite{JINA-REACLIB} recommended rates. The shaded bands indicate the 1$\sigma$ uncertainty, including statistical and systematic errors as well as contributions from the width-fluctuation treatment, and the possible contribution from direct capture.} 
 \label{fig:talyscalc_90Y}
 \end{center}
 \end{figure*}
We now use the obtained lower, middle and upper normalizations of the $\gamma$SF to constrain the input NLD and $\gamma$SF of $^{90}$Y using the nuclear reaction code TALYS-1.9~\cite{TALYS_18} to calculate the $^{89}$Y($n,\gamma$)$^{90}$Y cross section and astrophysical reaction rate. Specifically, we use the HFB+comb. NLD normalized with the parameters given in Table~\ref{tab:par1} to well reproduce the NLD data points \cite{goriely2008}. 
We also include the 30 first discrete levels of $^{90}$Y in the calculations.  
Further, we use the phenomenological, fitted models in Eq.~(\ref{eq:glofit}) as input with the parameters given in Table~\ref{tab:gsfpar}, including them as tabulated $E1$ and $M1$ strengths using the \textit{E1file} and \textit{M1file} keywords. 
For the neutron optical-model potential, we apply the one from Koning and Delaroche with
global parameters~\cite{koning03}. 
We consider the uncertainty in the treatment of the width fluctuations by using the default TALYS option (Moldauer, Refs.~\cite{moldauer,moldauer2} as well as the Hofmann-Richert-Tepel-Weidenm\"{u}ller model~\cite{hrtw,hrtw2,hrtw3}.
Further, we also take into account a possible contribution from direct capture as prescribed in Ref.~\cite{Xu2012}, using the TALYS keywords \textit{racap y} to invoke the direct-capture mechanism and \textit{ldmodelracap 2} to use total particle-hole state densities in the direct-capture calculation.
We propagate the errors quadratically as before to estimate $\approx 1\sigma$ uncertainties in the calculated cross section and the rate.
The resulting ($n,\gamma$) cross sections and reaction rates are shown in Fig.~\ref{fig:talyscalc_90Y}a and b, respectively. Note that using the constant-temperature (CT) NLD, $\rho_{CT}(E) = 1/T \exp{(E-E_0)/T}$~\cite{ericson1959,ericson1960} deduced from the $^{90}$Y data in Ref.~\cite{guttormsen_yttrium_2014}, gives a cross section and reaction rate very close to the middle normalization in this work. Hence, the HFB+comb. NLD and the CT NLD are fully compatible in this case.

We see from Fig.~\ref{fig:talyscalc_90Y}a that our upper limit best reproduces the data we compare with. 
This implies that the tentative value of $\left<\Gamma_\gamma\right> \sim 134$ meV
given in Ref.~\cite{mughabghab2006} is likely to be too low. Also, by looking at the two last individual radiative widths listed in Table~\ref{tab:gamma}, there could be an increasing trend. New measurements of both $\left<\Gamma_\gamma\right>$ and the ($n,\gamma$) cross section would be highly desirable to clarify the situation and provide higher precision.
As for the astrophysical rates shown in Fig.~\ref{fig:talyscalc_90Y}b, the BRUSLIB rate agrees 
rather well with our results for the middle normalization. The JINA REACLIB rate differs
significantly in shape between $T \approx 0.1-1$ GK, and the absolute value is also much higher
for $T\approx 4-10$ GK compared to the BRUSLIB one. Our current error band seems to capture both the library reaction rates except at the highest temperatures. 

\begin{figure}[!htb]
 \includegraphics[clip,width=.9\columnwidth]{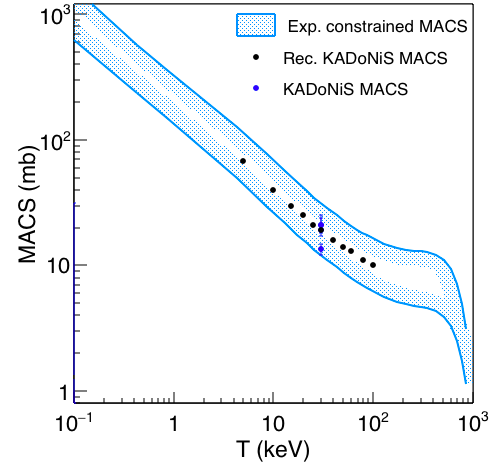}
 \caption{(Color online) Calculated $^{89}$Y($n,\gamma$)$^{90}$Y MACS compared with the experimental values compiled by KADoNiS at 30 keV \cite{sprocessN50,MAM78A, BAM77, AGM71} and for a range of temperatures according to Ref. \cite{bao2000}.} 
 \label{fig:talysmacs_90Y}
 \end{figure}

We have also calculated the Maxwellian-averaged cross section (MACS) and compared to experimentally available information compiled in the KADoNiS library \cite{kadonis}. The experimental results compiled in KADoNiS for 30 keV have statistical errors ranging from $3.2\% - 14.3 \%$ and the absolute value varies from 13.5-21 mb. The experimental MACS values are shown in Fig. \ref{fig:talysmacs_90Y} together with the present experimentally constrained MACS. Our results are in good agreement with the recommended KADoNiS values \cite{bao2000}. 
\section{Summary and outlook}
\label{sec:sum}
We have measured the $^{89}$Y($\gamma$,n) cross section between $S_n$ and $S_{2n}$ with high precision, providing a third data set that eventually could contribute to resolving the longstanding discrepancy between data from Livermore (Berman \textit{et al.}) and Saclay (Lepretre \textit{et al.}). Our errors are in the range of $\sim 3\% - 5\%$, where the larger relative error relates to the low cross section values measured close to $S_n$. The $^{89}$Y($\gamma$,n)-cross section measured in this work is rather consistent in shape with previous measurements, but our values are intermediate to the two previous measurements for $E_{\gamma}<$ 18 MeV. We are however unable to describe the details of the structure of the GDR by the phenomenological Generalized Lorentzian model without two components. It is known that ${89}$Y is a spherical system and this asymmetry can therefore not be contributed to deformation. 

We combined the $\gamma$SF obtained from $^{89}$Y(p,p$\gamma$) coincidence data and $E_{\gamma}<S_n$ with the $\gamma$SF obtained from the $^{89}$Y($\gamma$,n) and thereby describing experimentally most of the energy range 1.5 MeV $< E_{\gamma}<$ 20 MeV. The $\gamma$-ray strength function of $^{90}$Y for $E_{\gamma}<S_n$ has been studied using the Oslo method on $^{89}$Y(d,p$\gamma$) coincidence data. We assumed that structure effects can be neglected and that the $\gamma$SF of $^{89}$Y can be combined with that of $^{90}$Y to cover a large energy range. 

Our experimental results for $^{89}$Y and $^{90}$Y were combined into TALYS cross section and reaction rate calculations to constrain the $^{89}$Y($n,\gamma$)$^{90}$Y reaction cross section and the Maxwellian averaged reaction rate. In addition to the systematic uncertainty of the normalization, the gap in $E_{\gamma}$ where data is lacking also introduces substantial uncertainty in how to model the total $\gamma$-ray strength function. While our cross section results are consistent with several previous measurements and thus both BRUSLIB and JINA REACLIB, our systematic uncertainties stemming from the normalization parameters for the $\gamma$-ray strength function for $E_{\gamma}<8$ MeV are too large to be sensitive to the differences between the two reaction rate libraries in the temperature range of relevance for the s-process. Our MACS values are in good agreement with the recommended values of the KADoNiS library. 

While Hauser-Feshbach calculations of reaction cross sections cannot compete with experimental data, where available, the approach is needed in order to reliably predict for energy ranges and reactions where direct measurements are lacking. The $^{89}$Y($n,\gamma$)$^{90}$Y reaction cross section is vital in calculating the production of elements heavier than $A \sim 90$ in the s-process in stellar models, and has consequently been well studied experimentally (all thought not with small enough uncertainties for certain applications). Our experimentally based calculations demonstrate well the applicability of the approach of using experimental $\gamma$SFs and NLDs to constrain reaction cross sections, through the application of the Hauser-Feshbach formalism as implemented in TALYS, in this region of the nuclear chart. Future work will focus on obtaining experimental $\gamma$SFs and NLDs from particle-$\gamma$ coincidence data for unstable isotopes close to $N=50$ and using these results to constrain important cross sections for astrophysical applications. 

\acknowledgments
The authors wish to thank J.C.~M{\"{u}}ller, E.A.~Olsen, A.~Semchenkov and J.~Wikne at the Oslo Cyclotron Laboratory for providing excellent experimental conditions and T.~W.~Hagen, S.~Rose for taking shifts. 
Furthermore, the authors would like to thank H. Ohgaki of the Institute of Advanced Energy, Kyoto University, for making a large volume LaBr$_3$(Ce) detector available for the experiment at New SUBARU storage ring.
A.C.L. gratefully acknowledges funding through ERC-STG-2014, grant agreement no. 637686. A.C.L. would also like to thank A.~Koning for solving issues in the TALYS-1.9 release. 
G.M.T.  gratefully acknowledges funding of this research from the Research Council of Norway, Project Grant No. 262952. 
S.~G. is F.N.R.S. research associate. 
S.~S. acknowledges financial support by the Research Council of Norway, project grant no. 210007. 
M.~W. acknowledges support by the National Research Foundation of South Africa under grant no. 92789 and 83867. 
I.G. and D.F. acknowledge the support from the Extreme Light Infrastructure Nuclear Physics (ELI-NP) Phase II, project cofinanced by the Romanian Government and the European Union through the European Regional Development Fund - the Competitiveness Operational Programme (1/07.07.2016, COP, ID 1334). 
H.U. acknowledges the support from the Premier Project of the Konan University.
This work was partly supported by the IAEA and performed within the IAEA CRP on "Updating the Photonuclear data Library and generating a Reference Database for Photon Strength Functions" (F41032) and JPN-20564. 
A.V.V. acknowledges support from US Department of Energy DE-NA0002905. This work was partly performed under the auspices of the US Department of Energy DE-AC52-07NA27344 (LLNL) and DE-AC02-05CH11231 (LBNL). 
R.S. was supported by the European Commission within the
Seventh Framework Programme through Fission-2013-CHANDA (project no.605203).
\bibliographystyle{apsrev4-1}
\bibliography{ybibfile}
 
\vfill
\end{document}